\documentclass[twocolumn, twocolappendix]{aastex631}

\newcommand{\specialcell}[2][c]{%
\begin{tabular}[#1]{@{}c@{}}#2\end{tabular}}

\usepackage{subfigure}
\usepackage{graphicx}

\begin{document}


\title{You Shall Not Pass! The propagation of low/moderate powered jets through a turbulent interstellar medium}

\author[0000-0001-7522-1324]{Olga Borodina}
\affiliation{Center for Astrophysics $\vert$ Harvard \& Smithsonian, 60 Garden St, Cambridge, MA 02138, USA}

\author[0000-0001-7899-7195]{Yueying Ni}
\affiliation{Center for Astrophysics $\vert$ Harvard \& Smithsonian, 60 Garden St, Cambridge, MA 02138, USA}

\author[0000-0002-8573-2993]{Jake S. Bennett}
\affiliation{Center for Astrophysics $\vert$ Harvard \& Smithsonian, 60 Garden St, Cambridge, MA 02138, USA}

\author[0000-0001-6260-9709]{Rainer Weinberger}
\affiliation{Leibniz-Institut fur Astrophysik Potsdam (AIP), An der Sternwarte 16, 14482 Potsdam, Germany}

\author[0000-0003-2630-9228]{Greg L Bryan}
\affiliation{Department of Astronomy, Columbia University, 550 West 120th Street, New York, NY 10027, USA}

\author[0000-0002-3301-3321]{Michaela Hirschmann}
\affiliation{Institute of Physics, Laboratory for Galaxy Evolution, EPFL, Observatory de Sauverny, Chemin Pegasi 51, 1290 Versoix, Switzerland}

\author[0000-0002-5228-2244]{Marion Farcy}
\affiliation{Institute for Physics, Laboratory for Galaxy Evolution and Spectral Modelling, EPFL, Observatoire de Sauverny, Chemin Pegasi 51, 1290 Versoix, Switzerland}

\author[0000-0001-7271-7340]{Julie Hlavacek-Larrondo}
\affiliation{D\'{e}partement de Physique, Universit\'{e} de Montr\'{e}al, Succ. Centre-Ville,
Montr\'{e}al, Qu\'{e}bec, H3C 3J7, Canada}

\author{Lars Hernquist}
\affiliation{Center for Astrophysics $\vert$ Harvard \& Smithsonian, 60 Garden St, Cambridge, MA 02138, USA}



\begin{abstract}

Feedback from black hole-powered jets has been invoked in many cosmological simulations to regulate star formation and quench galaxies. Despite this, observational evidence of how jets might be able to affect their hosts remains scarce, especially for low power jets in halos smaller than clusters. Recent observations of outflows around FR0 galaxies, that host compact radio-loud sources, imply that lower-power jetted active galactic nuclei (AGN) may have a significant impact on their hosts through jet interactions with the interstellar medium (ISM). 
Using the \texttt{Arepo} code, we launch jets of low and intermediate power (10$^{38}$ --- 10$^{43}$\,erg\,s$^{-1}$) within a $\sim$kpc-scale periodic box with driven turbulence to study how the jets propagate through a turbulent ISM.
Our simulation results broadly fit into three different scenarios — jets penetrating easily through the ISM, becoming completely stalled, or the interesting intermediate stage, when jets are highly disturbed and redirected. 
We suggest that intermediate power jets do not have enough ram pressure to affect the turbulent structure of the ISM, and so only fill pre-existing cavities. Low-power jets are able to drive outflows in a hot phase ($>10^{4.4}$\,K). However, warm ($\sim\! 10^4$\,K) ionized gas outflows appear under certain conditions.
This work is part of the “Learning the Universe” collaboration, aiming to build next-generation cosmological simulations that incorporate a new prescription for AGN feedback.

\end{abstract}



\section{Introduction} \label{sec:intro}


Astrophysical feedback is one of the likely mechanisms that can lead to star formation quenching in galaxies \citep{Man2018}. The first main form of feedback is observable as galaxy-wide outflows caused by supernovae, i.e. stellar feedback \citep[e.g.][]{David1991, SpringelHernquist2003, Liu2023}. However, these outflows are not efficient enough for very massive galaxies --- the energy from stellar feedback may not be sufficient to overcome the binding energy of the dark matter halo \citep{Dubois2008, Su2019, Koutsouridou2022}.  The other key source of feedback, thought to dominate in high mass galaxies, is that from active galactic nuclei or AGN \citep{DiMatteo2005, Croton2006, Fabian2012}.

AGN are powered by the energy released during the accretion of magnetized gas onto a potentially spinning supermassive black hole (SMBH) in the centers of galaxies. AGN heat up the surrounding media and can source disk-driven winds \citep{Silk1998} and jets \citep{Blandford1982, Begelman1984}, which are usually seen in radiatively efficient and radiatively inefficient regimes respectively. The jets are highly collimated, fast-moving outflows of electrons and baryons \citep{Blandford2019}. They can propagate to scales from small (tens of parsecs) to large (hundreds of kiloparsecs) and can vary in power \citep[][and references therein]{Davis2020}.

The jets affect gas in and around the host galaxy. Observations show that AGN jets can be associated with significant outflows from the interstellar medium (ISM) \citep{Harrison2018}. Also, jets can heat hydrostatic atmospheres and offset cooling losses in galaxy clusters \citep{Binney1995, McNamara2012, Hlavacek-Larrondo2022}. 

AGN jets can be systematized based on spatial orientation, luminosity, spectral properties, and morphology \citep{Heckman2014}. If the flux density in the radio band is at least ten times higher than that in the optical, AGNs are termed radio-loud; otherwise, they are considered radio-quiet \citep{Kellermann1989}.
Historically, radio-loud AGN jets are classified depending on whether the radio brightness increases or decreases with distance from the SMBH; hosting galaxies were divided by \cite{Fanaroff1974} into two classes --- FRI and FRII.
Recently, a new class of radio-loud AGN, called FR0, has been proposed \citep{Baldi2016, Baldi2018}. Unlike FRI and FRII radio sources, FR0s are typically unresolved on scales of 1-3 kpc. 
Compared to other compact radio sources, FR0s have a flat radio spectrum, implying that the jet is powered constantly over millions of years \citep{Baldi2023}. Moreover, at a redshift $z < 0.5$, FR0s are at least 4 times more abundant than FRIs and FRIIs \citep{Baldi2023}. The prevalence of FR0s and the implication of constant powering suggests that we do not just see the early evolutionary stage of a FRI/II system, but a new class of objects where a jet is confined \citep{Baldi2018}. Moreover, FR0 galaxies, with their galaxy-scale jetted emissions, could play a crucial role in radio-mode feedback by potentially injecting energy into the ISM more efficiently than more powerful jets \citep{Baldi2023}.

Observations have also suggested that the interstellar medium (ISM) can affect jet propagation. For example, many compact jets have a very complicated asymmetric structure \citep{Kharb2019, Rao2023, Giovannini2023}. It has been suggested that the jet's interaction with the complex gas structures of the ISM plays an important role \citep{Fabbiano2022}. Indeed, as we observe in the Milky Way's center \citep{Heywood2022, Dinh2024}, the ISM is not uniform. Therefore, filaments and walls of dense gas may be able to deflect or stall low power jets.
Motivated by these recent discoveries,  theoretical studies are needed to determine why some jets do not propagate further and how they affect the ISM.

State-of-the-art cosmological simulations such as EAGLE, IllustrisTNG, and SIMBA \citep{Schaye2015, Pillepich2018, Dave2019} successfully match many observed galactic properties due to their AGN feedback implementations. For example, such simulations can reproduce the stellar content of massive galaxies \citep{Weinberger2018, Genel2018} or the bimodal distribution of galaxy colors \citep{Weinberger2017, Nelson2018}.
There are two commonly adopted types of feedback \citep{Churazov2005, Sijacki2007}. First, a high accretion rate mode, often implemented as a thermal energy injection, models the effect of radiative heating that corresponds to a `quasar mode' in observations \citep{King2015}. The second is kinetic feedback, typically happening at low accretion rates onto massive black holes, which aims to efficiently quench massive galaxies and reproduce the quiescent population through a `radio' mode \citep{McNamara2012}. 

The latter radio mode, however, is not truly a `jet' in cosmological simulations, since the central region cannot be resolved.
The physics involved in the launching and propagation of jets covers a huge range of length- and time-scales. As a result, it is infeasible to include jet-ISM interactions in large cosmological volumes. Therefore, AGN feedback implemented in cosmological simulations is based on heuristics and approximations, referred to as sub-grid models. For example, the threshold for switching from thermal to kinetic feedback in some simulations is dependent on the SMBH mass \citep{Weinberger2017}. 
Additionally, while cosmological simulations can reproduce some of the properties of hot bubbles driven by AGN observed in clusters \citep{HitomiCol2016, Prunier2024}, these bubbles are typically modeled in a simplistic manner. Specifically, in simulations, e.g. in IllustrisTNG and EAGLE, the interaction between the ISM and the jets is replaced by a powerful centralized injection of energy \citep{Weinberger2017, Schaye2015}.
These issues highlight the need to update the AGN feedback prescription, which is one of the key goals of the Learning the Universe\footnote{\url{http://learning-the-universe.org}} collaboration.

In order to better understand the mechanism by which jet feedback operates, galactic-scale jet-ISM studies can be conducted. 
Existing works have examined the evolution of jets in a non-uniform multiphase ISM \citep{Sutherland2007, Wagner2012, Mukherjee2016, Mukherjee2018, Cielo2018, Meenakshi2022, Dutta2024}. \cite{Wagner2012} and \cite{Dutta2024} showed that jet material fills hot gas channels whereas the jet head is slowed down by the clumpy medium. \cite{Mukherjee2016} found that high power jets escape the galactic center without affecting the ISM, whilst low power jets are trapped in the disk plane longer, which leads to further lateral spreading of the energy bubble.
However, many of these studies are based on the assumption that ISM structure is formed of small clumps, instead of filaments created by supersonic turbulence and efficient radiative cooling. Such filamentary structures are observed in the center of the Milky Way \citep{Heywood2022}. Moreover, existing studies do not fully explore the parameter space for jets with power lower than $10^{42}$\,erg\,s$^{-1}$, which are found to be abundant in observations \citep{Baldi2023}.

This work aims to model jet propagation in a more controlled turbulent environment shaped by driven turbulence and radiative cooling. We consider low and moderate power jets that would correspond to FR0 sources, and compare our results with jet propagation in a uniform medium in order to highlight the impact of the turbulent ISM structure. We also investigate whether low-power, stalled jets can affect the gas outside of the central kiloparsec.

We describe the details of the simulation setup of a turbulent medium and jet launching in Section \ref{sec:setup}. In Section \ref{sec:results} we describe the scenarios of jet propagation, as well as how jets affect the ISM. In Sections \ref{sec:discussion} and \ref{sec:conclusions} we summarize our results and discuss their implications for observations and cosmological simulations. 

\section{Simulation setup} \label{sec:setup}

We use the \texttt{Arepo} code developed by \cite{Springel2010} to run hydrodynamical simulations. \texttt{Arepo} solves the Euler equations using a finite volume approach on a quasi-Lagrangian, moving Voronoi mesh. Cells are refined and derefined to contain a specified target gas mass, except for the jet region, where additional refinement criteria are applied, as described in Section \ref{sec:jetsetup}. For simplicity, we ignore the effect of gravity in this setup, but will investigate this in future works.

\subsection{Setup of turbulent box} 

We set up a uniform box with a volume of (2\,kpc)$^3$, with 256$^3$ cells and target mass $M_t = 150\,\text{M}_\odot$. The box size is chosen to obtain a high resolution study of jet propagation on the ISM scale. Given the constraints imposed by the galactic disk height, the jet material is expected to escape into the halo beyond 1-2\,kpc.
In Appendix~\ref{sec:appendix-resolution}, we performed a resolution convergence test with 512$^3$ cells and found that it did not affect result in any notable differences.
We adopt periodic boundary conditions to facilitate turbulent driving. 
In order to mitigate non-physical effects due to periodic boundaries, we focus only on the central (1\,kpc)$^3$ and stop the simulation before the jet reaches the edge of the (2\,kpc)$^3$ box.
1\,kpc is also approximately the same scale at which FR0 jets are stalled and their central radio sources are not resolved \citep{Baldi2015}. In this work, the mean density $\bar{\rho} = 20$~cm$^{-3}$ of the central kiloparsec is chosen on the lower end of the typical electron number density of narrow line regions in AGN host galaxies at redshift $z<0.02$ \citep{Kakkad2018}. This value also represents the mean number density of the gas in the central kiloparsec of galaxies with stellar masses $10^{10} - 10^{11} M_\odot$ in the IllustrisTNG50 simulation \citep{Pillepich2019, Nelson2019}.

In order to simulate a realistic ISM, we drive solenoidal turbulence in the same way as \cite{Bauer2012}. 
In this work, turbulence is driven on the scale of 1\,kpc and then cascades down to smaller eddies. We add the physical scaling for specific energy rate $\epsilon$, using equations from \cite{Mocz2018}:

\begin{equation}
    \epsilon = \alpha \cdot \frac{c_s^3}{ 2^3 \cdot l},
    \label{eq:energy}
\end{equation}
where $c_s$ is the turbulence driving sound speed and $l$ is the turbulence driving scale. $\alpha$ is a parameter that allows us to  vary the strength of turbulence and thus the resulting Mach number of the turbulent medium.
We set the turbulence driving sound speed $c_s = 20$\,km\,s$^{-1}$, corresponding to a 10$^4$\,K ionized gas.



We subsequently define the Mach number with respect to this turbulent driving sound speed $c_s$:

\begin{equation}
    \mathfrak{M}=\frac{\langle v^2\rangle^{\frac{1}{2}}}{c_s} ,
    \label{eq:mach}
\end{equation}
where the square of the gas velocities in cells $v^2$ is averaged over the entire box.

Molecular gas in the galactic center can reach high values of Mach number $\mathfrak{M} \geq 30$ \citep{Henshaw2016}. If we assume that molecular and ionized gas have the same velocity dispersion, 
the respective Mach number for 10$^4$\,K gas is $\mathfrak{M} \geq 2$.
We set the fiducial Mach number $\mathfrak{M} \approx 4$, which can be obtained by setting $\alpha=1.0$ in Eq.~\ref{eq:energy}. To obtain Mach number $\mathfrak{M} \approx 2$ and $\mathfrak{M} \approx 8$ we used $\alpha=0.3$ and $\alpha=10.0$, respectively. 

We include radiative cooling for the gas \citep{Katz1996} and assume that the abundance is primordial and use a temperature floor of 10$^4$\,K for the cooling. In Appendix \ref{sec:appendix-metallicity}, we show that the results are not affected by including metal cooling.

In this work we do not include certain physics, such as relativistic effects and magnetic fields. For objects like FR0 jets at the scale of interest we can neglect general relativity \citep{Giovannini2023}. Considering that we launch low-power jets, the assumption of non-relativistic jets can be made. However, we plan to add a treatment of special relativity in future work. Some studies suggest that strong magnetic fields cause jet lobe distortion, while others argue that they prevent jet material from mixing with the ambient medium \citep[see][and references therein]{Bourne2023}.
Additionally, the magnetic field strengths around galactic centers are poorly constrained, making it speculative to include them as a parameter. Therefore, we focus on the hydrodynamic jet-ISM interaction first and defer the introduction of additional complexities (e.g. magnetic fields) to future work.

\begin{figure}[h!]
    \centering
    \includegraphics[width=8cm]{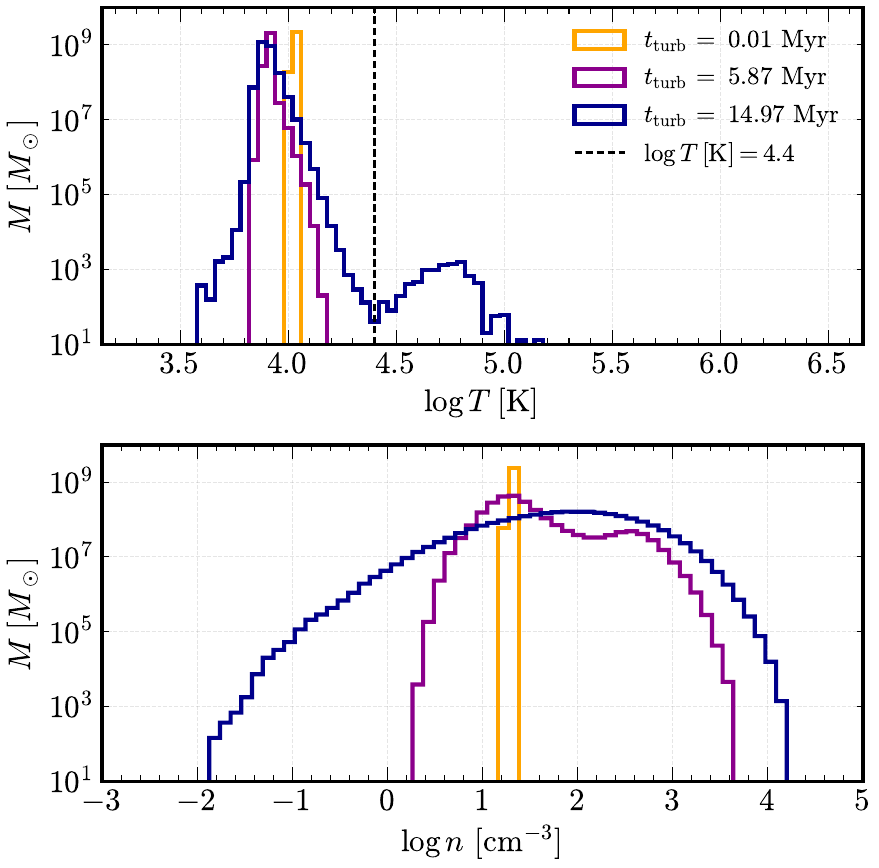}
    \caption{The top panel shows the temperature distribution of the cells weighted by their masses. In the beginning the gas in the box has the same temperature (yellow bins) and then develops into a multiphase medium (indigo bins). The black dashed line shows the temperature threshold we use to distinguish between gas phases. The bottom panel shows the number density distribution weighted by mass for the same snapshots as the panel above.}
    \label{fig:gas-phases}
\end{figure}

Initially, our box is isothermal with temperature $T_0 = 10^4$\,K, but turbulence and cooling lead to the development of two gas phases, which are apparent in the top panel of Fig.~\ref{fig:gas-phases}. Throughout this paper, we refer to the gas with temperature $\log T > 4.4$ as the ``hot'' phase, and gas below that threshold is considered ``warm''. We do not model low-temperature cooling and star formation in this work. We summarize the key parameter values of the turbulent box in Table~\ref{tab:parameters}.

\begin{deluxetable}{ccc}[tbh!]              \label{tab:parameters}
    \caption{Input parameters of the turbulent box}
\tablehead{
    \colhead{Variable} & \colhead{Description} & \colhead{Value}}
\startdata
        $\bar{\rho}$ & Mean number density & 20 cm$^{-3}$ \\
        $N_{\text{cell}}$ & Number of cells & $256^3$ \\
        $M_t$& Target mass & 150 M$_\odot$\\
         $c_s$& Turbulence driving sound speed & 20 km s$^{-1}$ \\
        $l$  & Turbulent driving scale & 1000 pc \\
         $\mathfrak{M}$ & Mach number & [2, 4, 8]\\
         $T_0$ & Initial uniform temperature & $10^4$ K\\
        $\gamma$ & Adiabatic index & $5/3$\\
\enddata
\end{deluxetable}

Due to the turbulence, the density distribution of the gas also changes.
In the bottom panel of Fig.~\ref{fig:gas-phases} we show how gas initially has a uniform density represented by yellow bins. After 15\,Myr both higher and lower density regions are created, so eventually the gas distribution follows a log-normal shape as expected analytically and numerically \citep{Padoan1997}.

We check if the turbulent medium reaches a steady state by seeing when the Mach number stabilizes. Analytically, we can estimate the timescale for the steady state regime using the largest eddy turnover time $t_\textrm{to} = \frac{L}{\mathfrak{M} \cdot c_s} \approx 12$\,Myr for Mach number $\mathfrak{M} = 4$. Once the ISM reaches a steady state, we start launching jets.

\subsection{Setting up the jet} \label{sec:jetsetup}

After we reach the steady state of the turbulent gas, at 15\,Myr for $\mathfrak{M} = [4, 8]$ and at 25\,Myr for $\mathfrak{M} = 2$, and while still driving turbulence, we launch jets with a constant power along the positive and negative x-axis, using the prescription described in \cite{Weinberger2017b, Weinberger2023}. We set the black hole smoothing length to be 90\,pc, meaning that an inner sphere with radius $r = 30$\,pc becomes the jet injection region. The outer spherical shell with $30\, \text{pc} < r < 90 \,\text{pc}$ is used to estimate the properties of the surrounding gas. Part of the jet energy is first deposited thermally to ensure pressure equilibrium between the jet injection region and the surrounding material. Then, the rest of the energy (the dominant part) is deposited kinetically. We set the jet density within the inner sphere as $10^{-26}\text{ g} \text{ cm}^{-3}$. The opening angle is set to 0 degrees, directing the injected momentum entirely along the x-axis. This approximation is reasonable for small-scale jets \citep{Begelman1984}.

To track jet propagation, we add a tracer scalar $X_\mathrm{jet}$, defined as the mass fraction of jet material in a cell. The jet tracer scalar value inside the jet injection region is set to $X_\mathrm{jet}= 1$ and propagates with the mass flux to other cells. We consider material to be a part of the jet if $X_\mathrm{jet} > 10^{-3}$.

To better resolve jet propagation in the jet region, we add additional refinement of the jet material. As we list in Table~\ref{tab:parameters}, the target mass for the ambient medium is still set to $M_{\rm t} = 150 \, \text{M}_\odot$. Cells that are flagged as jet material are refined to the target volume $V_t = 729\,\textrm{pc}^{3}$, corresponding to a cell size of $9$\,pc if cells had cubic form. The parameters that we use to set up the jet are summarized in Table \ref{tab:parameters-jet}.

\begin{deluxetable}{ccc}[tbh!]
\caption{Jet parameters}
\label{tab:parameters-jet}
\tablehead{
    \colhead{Variable} & \colhead{Description} & \colhead{Value}}
\startdata
        $R_\mathrm{hsml}$ & Black hole smoothing length & 90\,pc \\
        $\rho_\mathrm{jet}$ & Black hole jet density &  $10^{-26}\,\text{g}\,\text{cm}^{-3}$ \\
        $V_{\rm t}$ & Target volume & 729\,pc$^{3}$ \\
        $X_\mathrm{jet, min}$ & \specialcell[t]{Mass fraction threshold \\ for jet refinement}  & $10^{-3}$ \\
        $\theta$ &Opening angle & $0^{\circ}$ \\
        $t_0$ & Switch on time & 
        \specialcell[t]{15\,Myr for $\mathfrak{M} = [4, 8]$ \\ 25\,Myr for $\mathfrak{M} = 2$}
        \\
        $L_{\text{jet}}$ & Jet powers & $10^{[38, 40, 43]}$\,erg\,s$^{-1}$ \\
\enddata
\end{deluxetable}




\section{Results} \label{sec:results}

In this section, the results of the simulation suite are presented. For subsections \ref{sec:jet-scenarios} to \ref{sec:outflows} we consider jets with a constant power of $[10^{38}, 10^{40}, 10^{43}]$\,erg\,s$^{-1}$ that are launched in boxes filled with identical gas structure with Mach number $\mathfrak{M} = 4$. In subsection \ref{sec:dif-mach} we present results of a $10^{40}$\,erg\,s$^{-1}$ jet propagating in different turbulent structures with Mach numbers $\mathfrak{M} = [2,4,8]$.

\subsection{Jet propagation scenarios}
\label{sec:jet-scenarios}

\begin{figure*}[tbh!]
    \centering
    \includegraphics[width=0.75\textwidth]{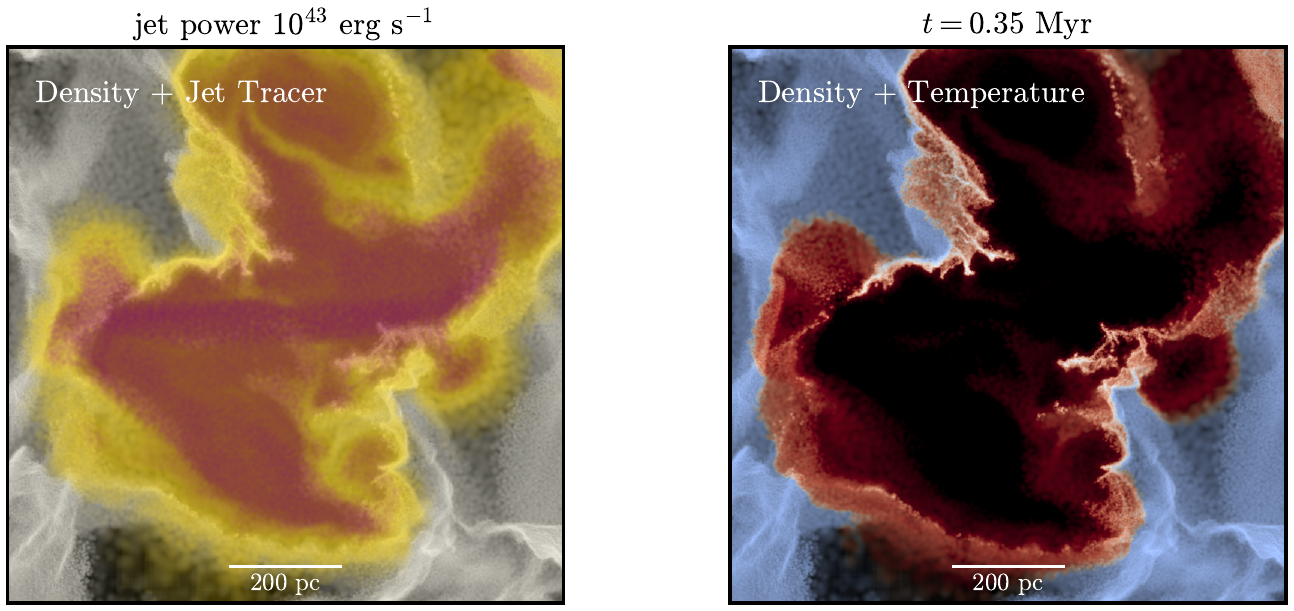}
    \includegraphics[width=0.75\textwidth]{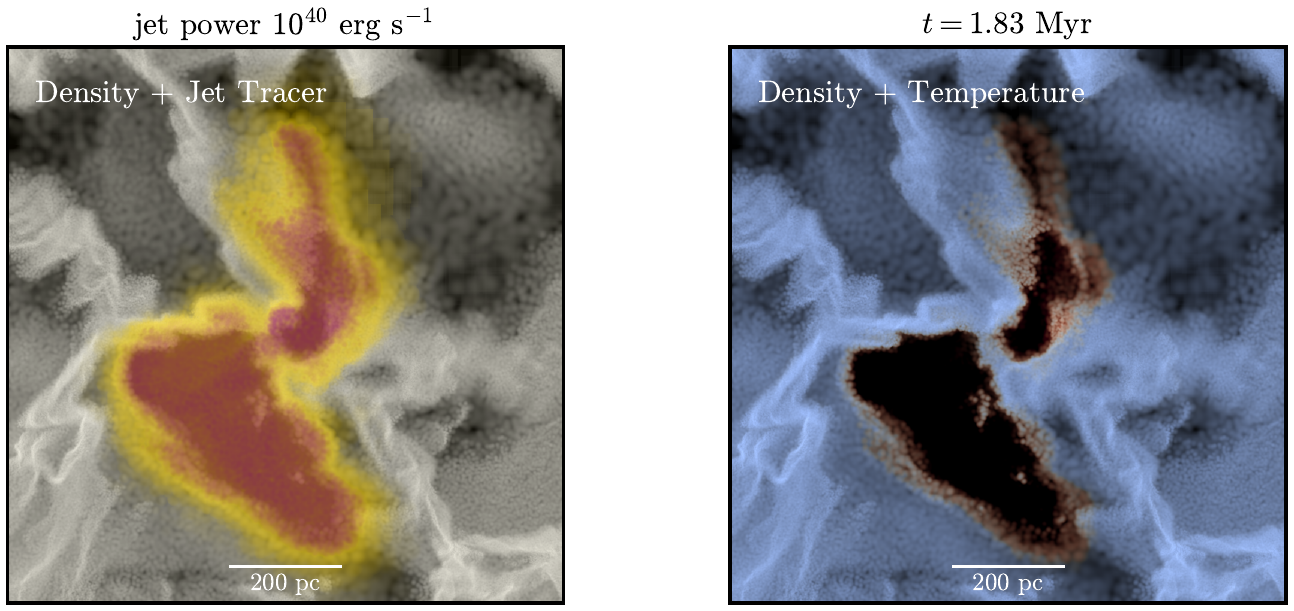}
    \includegraphics[width=0.75\textwidth]{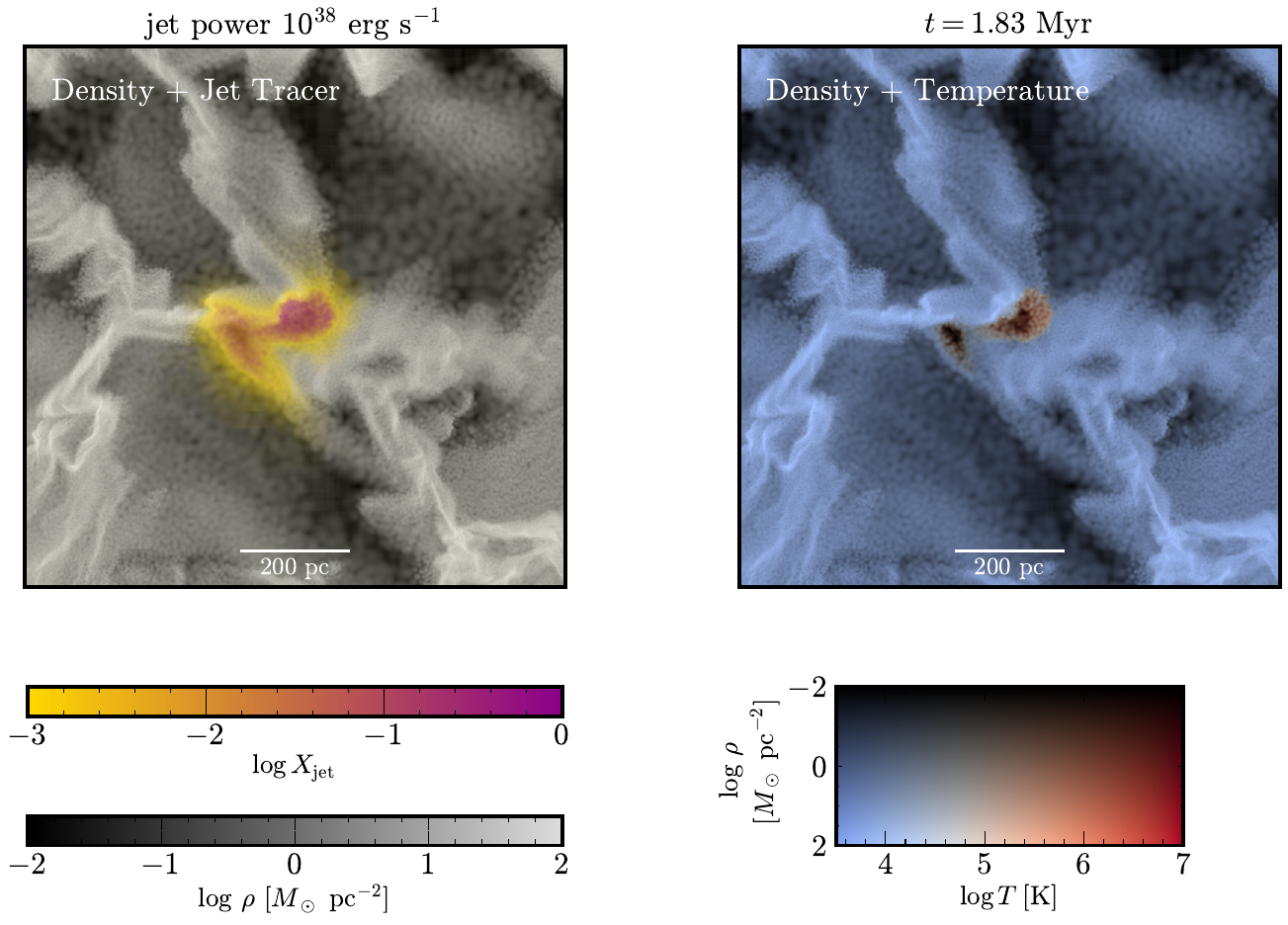}
    \caption{Jet propagation projections for jet powers of $10^{43}$ (top),  $10^{40}$ (middle), and $10^{38}$\,erg\,s$^{-1}$ (bottom) in the turbulent box with Mach number $\mathfrak{M} = 4$. The left-hand panels show black-white density projections with overlay of jet tracer scalar. Right-hand panels show density maps colored by temperature value. Projection maps have the depth of 100\,pc. }
    \label{fig:projection-maps}
\end{figure*}

In Fig.~\ref{fig:projection-maps} we show projection maps of gas density centered at the BH with a depth of 100\,pc overplotted with the jet tracer scalar value (left column) and colored by gas temperature (right column). For intermediate and low-power jets we show the jet propagation at the same time, approximately 1.83\,Myr after the jet is launched, to clearly show the difference between these runs. However, the strong jet with power $10^{43}$\,erg\,s$^{-1}$ reaches the box boundary within the first million years, therefore we choose an earlier snapshot.

Powerful jets ($\gtrsim 10^{43}$\,erg\,s$^{-1}$) can typically easily penetrate through turbulent environments (top panels in Fig.~\ref{fig:projection-maps}). Interestingly, even though most of the jet material moves along the launching direction, it also expands laterally. The collimated jet is shown by a pink-colored horizontal region, but jet material is shown at the same distance as the jet head in nearly all direction (see the top left panel of Fig.~\ref{fig:projection-maps}). 

Less powerful jets, with a power of $10^{40}$\,erg\,s$^{-1}$, are disrupted by turbulence, which causes the jet material to form bubble-like shapes and bend. See the middle panels in Fig.~\ref{fig:projection-maps} for an illustration of such a scenario. Where turbulence creates a low-density region near the jet injection region, the jet can fill the cavity -- but the jet cannot penetrate far in the launch direction. Despite this, jet material can still reach the box boundary, but notably \textit{not} in the original jet launch direction. 

In a third scenario, with a jet power of $10^{38}\,$erg\,s$^{-1}$, the jet is completely stopped within the first kiloparsec (bottom panels of Fig.~\ref{fig:projection-maps}). We re-emphasize that the middle and bottom panels show the box at the same snapshot, with the same turbulent structure in place. We can see though how difficult it is for the lowest power jet to penetrate a high-density area to fill the cavity to the left, unlike the jet with power $10^{40}$\,erg\,s$^{-1}$ which fills this cavity with hot jet material.


To determine the impact of ISM turbulence on the jets, we also launch them in two different uniform, turbulence-free boxes. The first has the same \textit{average} gas number density as the turbulent box. In large cosmological simulations, we do not generally have information about the kpc-scale ISM structure as it is often smoothed artificially by assuming an effective equation of state for dense gas \citep{SpringelHernquist2003}. Thus, the average number density of the turbulent box is representative of the density in centers of galaxies in the IllustrisTNG50 simulation \cite{Pillepich2019, Nelson2019}. The second modeled uniform box has dilute gas, with a density corresponding to the 10th percentile of the turbulent box cells' number density ($0.1\,\text{cm}^{-3}$). This test shows the impact of the multiphase \textit{structure} of the ISM -- high-density walls and filaments -- on jet propagation. An example of jets launched in dilute gas is shown in the Appendix \ref{sec:appendix-uniform}.

Fig.~\ref{fig:jet-propagation} quantifies the jet propagation over time.
For different radii of a sphere, centered on the black hole, we calculate the jet mass contained in it and divide by the total jet mass in the box. In Fig. \ref{fig:jet-propagation}, we plot the resulting fraction with shading and separately highlight the radius $r_{80}$, which contains 80\% of the jet mass.

\begin{figure*}[htb!]
    \centering
    \includegraphics[width=\textwidth]{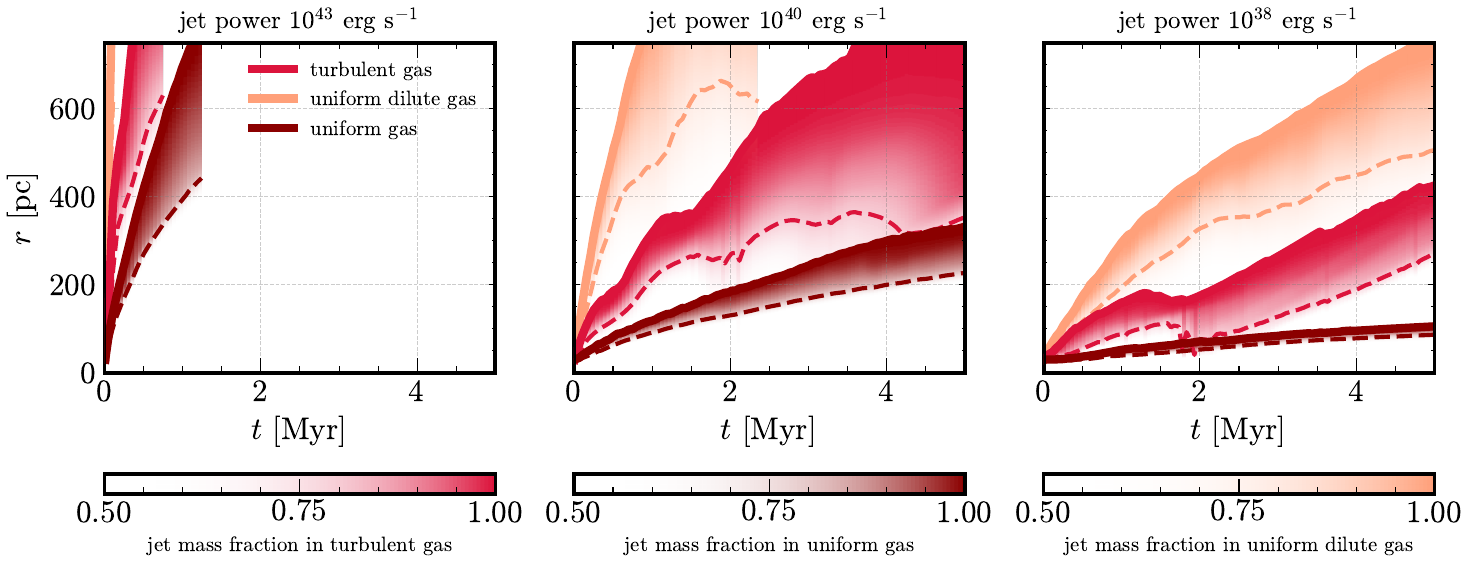}
    \caption{Evolution of the jet material propagation as a function of time for three different jet powers --- $10^{43}$\,erg\,s$^{-1}$ in the left panel, $10^{40}$\,erg\,s$^{-1}$ in the middle panel, and $10^{38}$\,erg\,s$^{-1}$ in the right panel, all in a turbulent box with Mach number $\mathfrak{M} = 4$. The shading represents the fraction of the jet material mass contained within the respective distance. The upper boundary of the shading indicates the furthest distance the jet has propagated. The dashed line corresponds to $r_{80}$.
    }
    \label{fig:jet-propagation}
\end{figure*}

For the run with jet power $10^{43}$\,erg\,s$^{-1}$ (see the left panel in Fig.~\ref{fig:jet-propagation}) we show that the jet propagates in the turbulent box more slowly than in the dilute uniform medium. Equivalently, propagation in the turbulent box is faster than in the uniform medium with the same mean number density. But in the end, a powerful jet escapes the central kiloparsec no matter the state of the medium. As shown in projection maps in Fig.~\ref{fig:projection-maps}, a high-power jet in a turbulent medium tends to heat up the gas in all directions. Unlike this, jets in a uniform medium create elongated cocoons with little propagation laterally (see Fig. \ref{fig:projection-map-uniform}).

In contrast, lower power jets are significantly affected by the turbulence. Most of the mass of the jet with power $10^{40}$\,erg\,s$^{-1}$ (middle panel Fig.~\ref{fig:projection-maps}) is confined in the central kiloparsec, but at least some material leaves this region. It is shown by the divergence between the maximum and median distances of the jet material. 
The jet with a power of $10^{38}$\,erg\,s$^{-1}$ does not have enough momentum to create a channel to expand into and therefore is completely stalled (see right panel Fig.~\ref{fig:projection-maps}).

In Fig.~\ref{fig:jet-propagation} the dark red dashed line that represents jet propagation in the uniform box of the same mean density has a different behavior from the solid line that shows jet propagation in a turbulent medium. First, unlike jets in a realistic medium, jets monotonically expand in the uniform box. Second, for each jet power, we observe jet material rising further in the turbulent medium which is shown by the top line of the filled regions.

Jets also propagate differently in a turbulent medium compared to the dilute uniform one as well. If we look at the light pink diagonally hatched regions in Fig.~\ref{fig:jet-propagation}, that represent jet propagation in the dilute uniform gas, we can observe that jets always propagate further and faster than in a turbulent medium. Additionally, compared to the monotonic expansion in uniform boxes, the jets in turbulent media do have a boost in their propagation when there is a cavity created by the turbulence.

Our results show how small-scale turbulent structure can significantly impact the evolution of jets. Therefore, in future studies we should account for such small scale, multi-phase gas structures. In the next section, we investigate the physical processes that lead to the jet propagation scenarios we have outlined here.

\subsection{How are jets stalled?}



The light but fast-moving jet material moves through the background medium, interacting hydrodynamically. In a low density uniform medium, the jet can push through the background material essentially unhindered. However, as the turbulence leads to the creation of high-density walls with low-density cavities between them, jet propagation becomes more complicated. Jets can easily propagate through the latter, following the path of the least resistance. Then, when the jet encounters a wall or filament, it may be stalled or redirected if it lacks sufficient ram pressure.

We calculate ram pressure for each cell using the following formula:

\begin{equation}
    p_{\mathrm{ram}, i} = \rho_i \cdot \sum_j{v_{ij}^2},
\end{equation}
where $v_{ij}$ is the j-th component of velocity of the i-th cell with mass density $\rho_i$.

\begin{figure}[htb!]
    \centering
    \includegraphics[width=\linewidth]{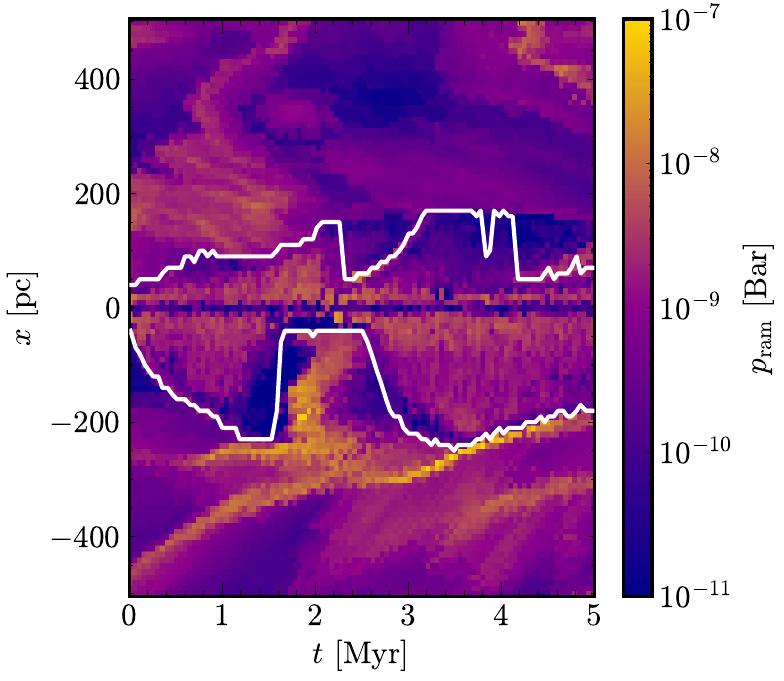}
    \caption{The white lines represent the furthest position of jet cells along the $x$-axis to the left ($x < 0$) and to the right ($x > 0$) as a function of time for the simulation with $\mathfrak{M} = 4$ and jet power $10^{40}$~erg\,s$^{-1}$. The color map shows the ram pressure values along the $x$-axis at each time.}
    \label{fig:ram-pressure}
\end{figure}

To study the evolution of ram pressure, we take cells contained within a cylinder of radius 30 pc along the $x$-axis (at $y = z = 0$) and calculate their mass-weighted average ram pressure as a function of $x$. We do this for each snapshot and plot these values in Fig.~\ref{fig:ram-pressure} to highlight the changing structure of the turbulent ISM. 

Bright yellow regions show the location of the dense walls. The white lines indicate the locations where the jet material mass fraction \textit{first} drops below $X_\textrm{jet} = 2 \cdot 10^{-3}$ to the left and to the right from the jet injection region. Fig.~\ref{fig:ram-pressure} shows that the extent of the jet propagation is highly correlated with the ISM structure. For example, up until 1.5\,Myr, the jet's ram pressure (region between the white lines in Fig.~\ref{fig:ram-pressure}) is higher than that of the ISM (region outside the white lines in Fig.~\ref{fig:ram-pressure}). Therefore, the jet propagates up until it hits the ``brighter'' zone with higher ram pressure. In order to cut through the walls, we would therefore need a jet with sufficiently high ram pressure --- requiring either a higher velocity from a higher jet power or a ``heavier'' jet with a higher mass loading. 

\subsection{Analytical model}

We propose a simple model to determine when the jet is able to push through the walls of the turbulent ISM and so write down the jet properties that are required to escape the ISM. We do this by comparing momentum fluxes, or ram pressures of the jet and ISM.  We begin by noting that the jet ram pressure is given by
\begin{equation}
p_{\rm jet} = \rho_{\rm jet} v_{\rm jet}^2 = \left( \frac{\rho_{\rm jet}^{1/2} L_{\text{jet}} }{A} \right)^{2/3},
\end{equation}
where we assume the jet velocity $v_{\rm jet} = (L_{\text{jet}}/A \rho_{\rm jet})^{1/3}$ is constant as the jet remains collimated with cross-sectional area $A$.

The gas density distribution in an isothermal turbulent ISM 
is well-described by a log-normal distribution \citep{Padoan1997}. The typical high densities are approximately $(1 + b \mathfrak{M}) \bar{\rho}$, where $b$ is a constant that depends on the nature of the turbulence (solenoidal or compressional) \citep{Federrath2008}. We assume a mix of both such that $b \approx 0.5$. The gas velocity distribution is assumed to be Gaussian with no correlation to the density, so typical velocities are $\mathfrak{M} c_s$. Taken together, this implies a typical ISM ram pressure of
\begin{equation}
p_{\rm ISM} \approx  (1 + b \mathfrak{M}) \mathfrak{M}^2 \bar{\rho} c_s^2.
\end{equation}

Interestingly, the peak of the ISM momentum flux distribution is a strong function of the Mach number, in agreement with the trends seen in Figures \ref{fig:ram-pressure} and \ref{fig:ram-pressure-mach}.

Equating these two ram pressures, we can derive an expression for the minimum jet luminosity required to break out of the ISM
\begin{equation}
L_{\rm jet,min} = \bar{\rho}^{3/2} c_s^3 (1 + b \mathfrak{M})^{3/2} \mathfrak{M}^3 A \rho_{\rm jet}^{-1/2}.
\end{equation}

Evaluating this for our typical ISM parameters for the $\mathfrak{M} = 4$ case, we find $L_{\rm jet,min} = 2 \times 10^{41}$ erg/s. This is in rough agreement with our findings that only the highest jet luminosity can escape the (1\,kpc)$^3$ box, although the model is relatively simple so more careful calibration is required. Finally, we note the inverse dependence on the jet density for fixed jet luminosity. This occurs because it is the jet momentum, not energy, which is responsible for pushing the jet working head further into the ambient medium. As noted earlier, if the jet remains collimated, this scaling is correct; however, if the jet is launched with an opening angle or if the cocoon effectively expands due to lateral expansion (see the model and simulations in \citealt{Su2021}), then these scalings would change, generally becoming more difficult to escape the ISM due to the lower ram pressure of the cocoon leading edge. Finally, we remark that we have assumed rapid cooling -- if the jet energy is not rapidly lost, then the expansion may become more spherical and the scalings will again be different.

\subsection{Implications for galactic scales}
\label{sec:outflows}


%

While many simulations have examined the impact of jets on massive galaxy clusters \citep[e.g.][]{Bourne2017, Weinberger2023} agreeing qualitatively on their feedback effect, the role of jets in galactic environment is less clear \citep{Gaibler2012, Mukherjee2018, Talbot2022}.
Observations regularly detect outflows even in galaxies with compact, low-power AGN sources \citep{Cheung2016, Roy2021}. Therefore, we investigate whether such jets can drive gas outflows (replicating ejective AGN feedback) through their interaction with the ISM.

We calculate the radial velocity of the gas cells at the same time in the turbulent box with and without jets, allowing us to compare the impact that jets have on the ISM. We choose the cells that lie in a spherical shell between 400 and 500 pc, since for jet power values of $10^{38}$ and $10^{40}$\,erg\,s$^{-1}$ the majority of jet material is located within 400pc. Therefore, we can investigate whether stalled jets can affect a galaxy outside of the very central region. We start measuring outflows when the jet material approaches $r_{100} \approx 400$\,pc or reaches its maximum distance. Then, we average the mass in each velocity bin over the following 1\,Myr in order to mitigate the random flow of the turbulence.

\begin{figure*}[tbh!]
    \centering
    \includegraphics[width=\textwidth]{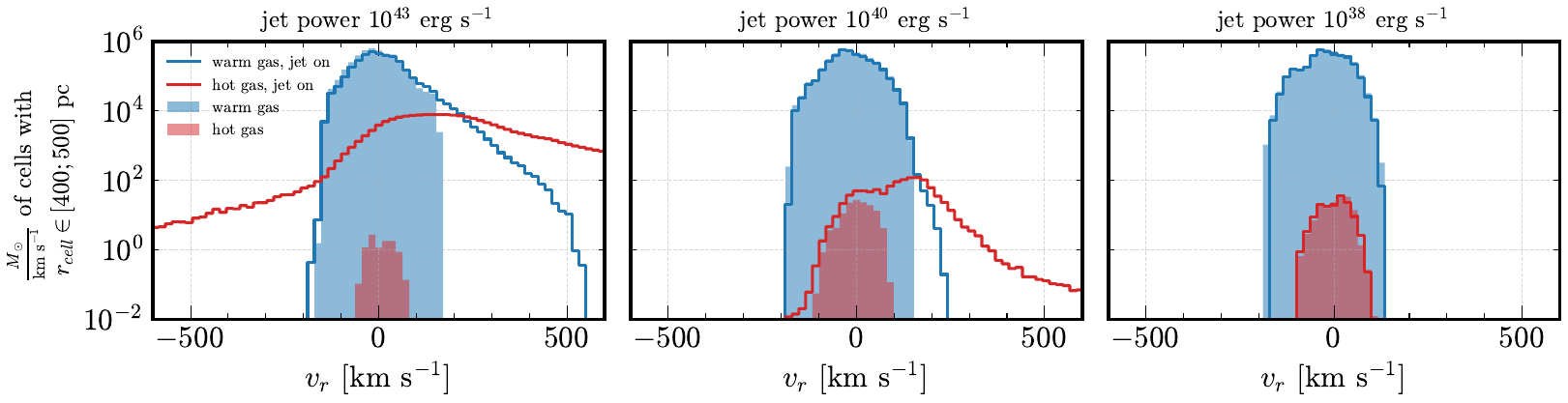}
    \caption{Mass-weighted radial velocity distribution of the gas phases for different jet powers --- $10^{43}$ (left),  $10^{40}$ (middle), and $10^{38}$\,erg\,s$^{-1}$ (right) averaged over 1\,Myr. Blue and red unfilled histograms represent the radial velocity distributions for the warm and hot phases, respectively, with the jet on. Blue and red filled bins show the radial velocity distribution of the warm and hot gas phases, respectively, in the turbulent box without jets, for the same time as in the boxes with jets. Each velocity bin is averaged over 1\,Myr --- for $10^{43}$\,erg\,s$^{-1}$ from  0 to 1\,Myr, for  $10^{40}$\,erg\,s$^{-1}$ from 1.5 to 2.5\,Myr, and for $10^{38}$\,erg\,s$^{-1}$ from 3.5 to 4.5\,Myr.}
    \label{fig:outflows}
\end{figure*}

Fig.~\ref{fig:outflows} shows that before we turn on the jet, gas has an initially symmetric radial velocity distribution due to the turbulence. We note that the initial distribution is the same for all three panels since we start jets in identical boxes.

After we turn on the jet, the outflow picture in both phases changes depending on the jet's power. If we launch a very weak jet, with jet power $10^{38}$\,erg\,s$^{-1}$, the velocity structure of the box remains unaffected (see the right panel in Fig.~\ref{fig:outflows}). We start to observe outflows, with velocities higher than the turbulence, in both warm and hot phases for the intermediate jet power (see the middle panel in Fig.~\ref{fig:outflows}). From the histogram, we see that hot gas outflows have velocities lower than 750 km\,s$^{-1}$. There are also outflows in the warm phase reaching a velocity of 250\,km s$^{-1}$. For the most powerful jet in our study (left panel \ref{fig:outflows}) we detect strong outflows in the hot phase (velocities of 0.1$c$). We also observe outflowing warm gas with a velocity of up to 500 km\,s$^{-1}$. For the highest power jet negative radial velocity bins show backflow material which is also found in the simulations of jets propagating in the uniform media \citep{Cielo2017}. 
Even though we find hot gas outflows to be more energetic, they are hard to estimate from observations.

Outflows in uniform media also differ from those caused by jets launched in turbulent media. For a jet power of $10^{43}$\,erg\,s$^{-1}$, strong hot outflows are present in a dilute medium, with low backflow, but there is a negligible amount of warm, slow outflows.  In a dense medium, strong jets produce slower hot outflows; however, the warm outflows and hot backflows are similar to those observed in turbulent media. For lower-power jets, outflows reaching the 400–500\,pc region are observed only in the box with a dilute medium, where weak and slow outflows occur in the hot phase. In the case of jets with a power of $10^{38}$\,erg\,s$^{-1}$  these outflows have velocities lower than the turbulent background, making them undetectable in Fig.~\ref{fig:outflows}.

\subsection{Differences in turbulent structure}
\label{sec:dif-mach}

 Our results show that a turbulent ISM affects jet propagation, with the positions of walls and cavities determining the morphology of the jet. To test how the stochasticity of turbulent structure affects jet evolution, we perform tests with boxes of the same Mach number but with different turbulent structures, realized through using different jet starting times. We observe that despite variations in the propagation of individual jets, the general trends discussed in our results are independent of the initial conditions of the turbulent medium (see Appendix~\ref{sec:appendix-stochasticity}).
 
 We also studied how jet propagation changes depending on different turbulence strengths. We create two more turbulent boxes with Mach numbers $\mathfrak{M} = 2$ and $\mathfrak{M} = 8$ by changing the turbulence driving energy of equation \ref{eq:energy}. The box with $\mathfrak{M} = 2$ contains larger cavities that evolve more slowly, whereas in a box with higher Mach number, $\mathfrak{M} = 8$, cavities are smaller and have shorter lifetimes (see Fig.~\ref{fig:turb-machs}). Then we launch a jet with a power of $10^{40}$\,erg\,s$^{-1}$ in the same way as previously. For Mach number $\mathfrak{M} = 2$ the largest eddy turnover time is $t_\textrm{to} = 24$\,Myr, therefore we launch the jet at 25\,Myr instead of 15\,Myr for $\mathfrak{M} = [4, 8]$.
 
\begin{figure*}[htb!]
    \centering
    \includegraphics[width=0.9\textwidth]{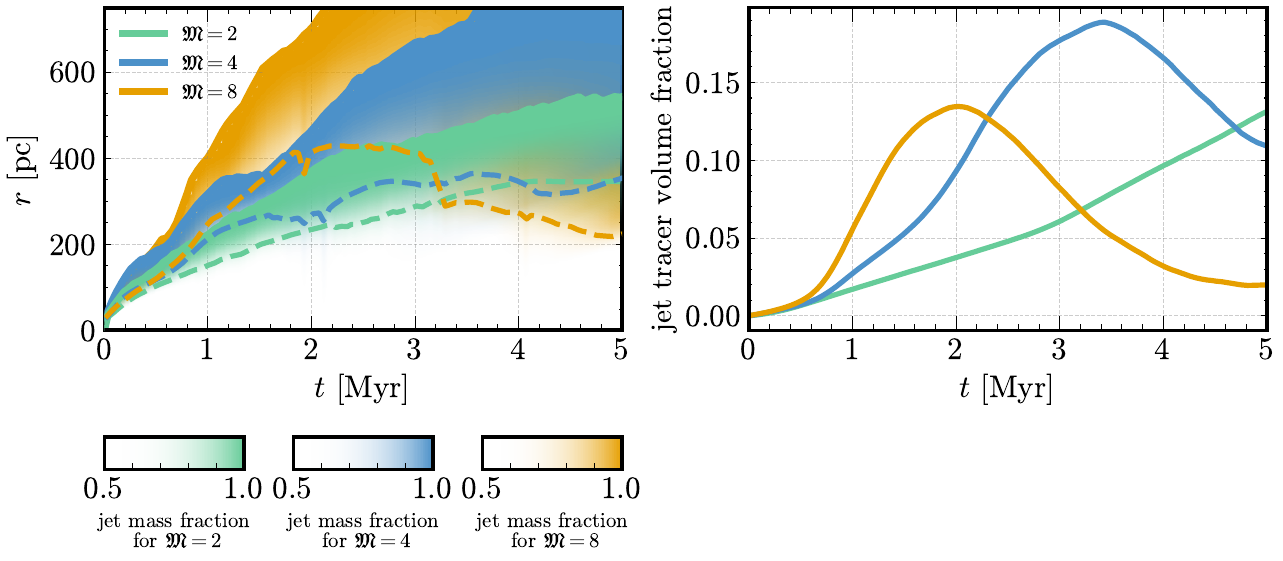}
    \caption{The left panel shows the evolution of jet material mass fraction
    as a function of time in boxes with three different Mach numbers $\mathfrak{M} = 2, 4, 8$.
    As in Fig.~\ref{fig:jet-propagation}, the shading corresponds to the fraction of jet material contained within the respective radius. The dashed line represents $r_{80}$ for each turbulent box. The right panel shows the jet tracer volume fraction within the central kiloparsec as a function of time for jet propagating in the same three turbulent boxes.}
    \label{fig:jet-propagation-mach}
\end{figure*}

Then we repeat the same analysis we performed for the previous runs. First, we measure how far the jet propagates. Besides the time evolution of the jet mass fraction (left panel of Fig.~\ref{fig:jet-propagation-mach}), we also plot the fraction of the central (1\,kpc)$^3$ volume occupied by cells with jet material, i.e. with $X_\mathrm{jet} > 10^{-3}$(right panel of Fig.~\ref{fig:jet-propagation-mach}).

In the left panel of Fig.~\ref{fig:jet-propagation-mach}, we show that the higher the Mach number of the ISM, the higher the slope of the top part of the shaded region. This means that stronger turbulence helps to propagate \textit{some} jet material further out to large distances, in the form of filled cavities. However, most of the jet material is stalled for the box with higher Mach number (see yellow line in left panel of Fig.~\ref{fig:jet-propagation-mach}) while jets in the ISM with lower Mach number steadily propagate further out. 

The right panel of Fig.~\ref{fig:jet-propagation-mach} shows more clearly that in the ISM with the highest Mach number, the jet never occupies much volume. The decline in volume fraction after 2\,Myr following jet launch for $\mathfrak{M} = 8$ and after 3.5\,Myr for $\mathfrak{M} = 4$ is explained by material leaving the central kpc as filled cavities move outwards, without being replaced by newer ``bubbles''.

\begin{figure*}[htb!]
    \centering
    \includegraphics[width=0.8\textwidth]{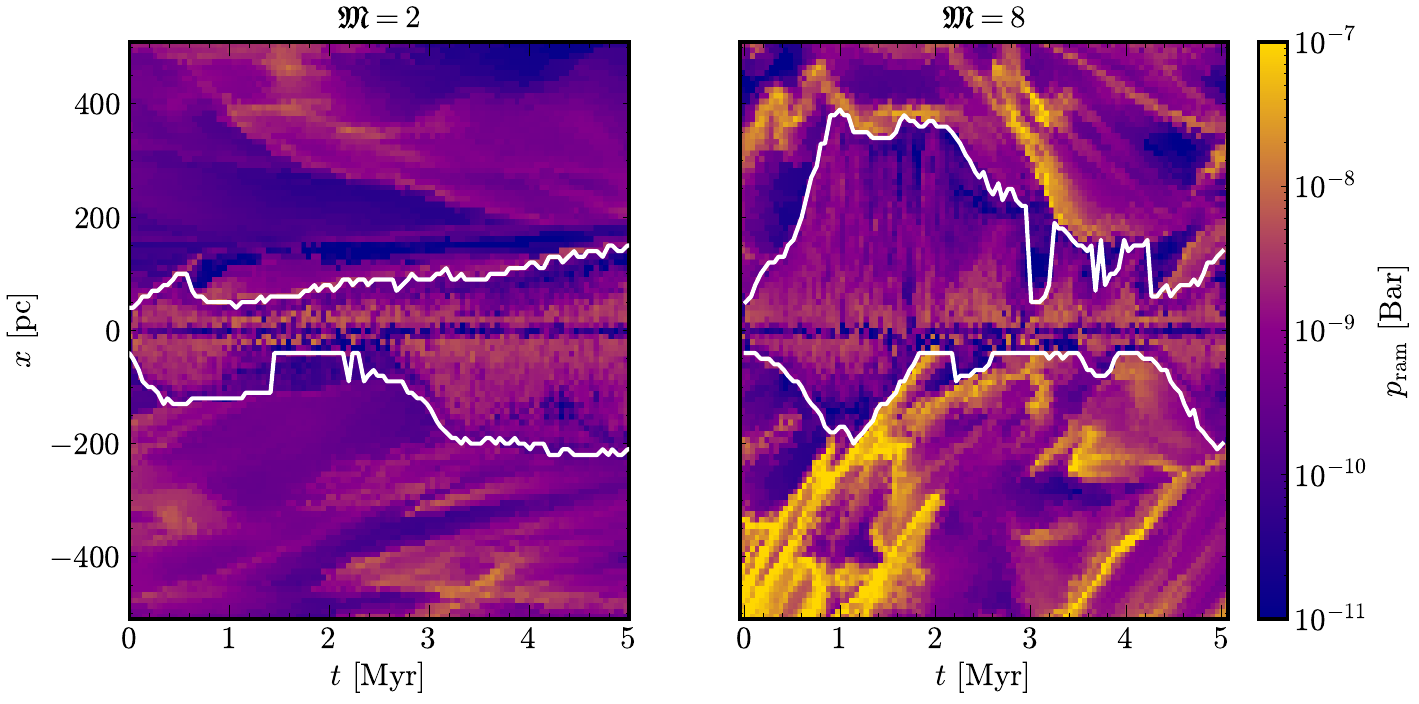}
    \caption{The white line represents the furthest position of jet cells along the $x$-axis as a function of time. The color map shows the mass-weighted average ram pressure values in a cylinder along the $x$-axis as a function of time. The left panel shows the results for $\mathfrak{M} = 2$, the right panel shows $\mathfrak{M} = 8$.}
    \label{fig:ram-pressure-mach}
\end{figure*}

We can clearly see the significance of turbulent structure on the ram pressure diagram (Fig.~\ref{fig:ram-pressure-mach}). As also shown in Fig.~\ref{fig:ram-pressure}, the jet propagates as long as its ram pressure is higher than that of the ISM. Therefore the walls that are denser and more frequently in the path of the jet in the $\mathfrak{M}=8$ run (right panel) more strongly affect jet propagation.
The walls in the box with lower turbulence (left panel) have lower peak density and also move more slowly compared to the right panel. We see that the ISM structure for  $\mathfrak{M} = 2$ has fewer low-density cavities, which leads to a lack of regions that the jet can easily fill.

\begin{figure*}[tbh!]
    \centering
    \includegraphics[width=\textwidth]{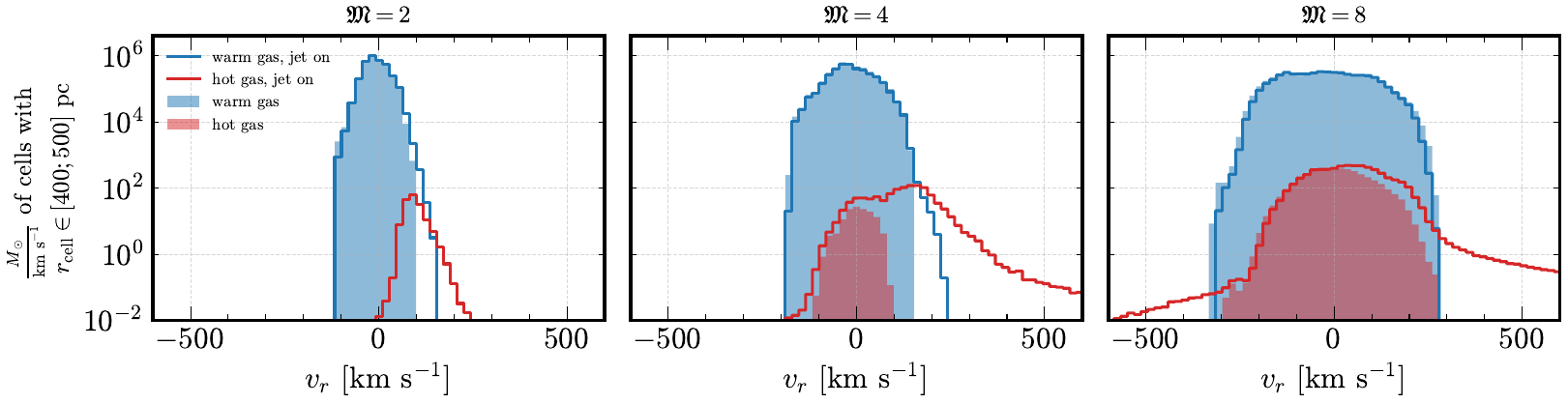}
    \caption{Mass-weighted radial velocity distribution of the gas phases for boxes with different Mach numbers $\mathfrak{M} = 2, 4, 8$. Blue and red unfilled histograms represent the radial velocity distributions for the warm and hot phases, respectively, with the jet launched. Blue and red filled bins show the radial velocity distribution of the warm and hot gas phases, respectively, in the turbulent box for the same time as in the boxes with jets. Each velocity bin is averaged over 1\,Myr --- for $\mathfrak{M} = 2$ from  2.5 to 3.5\,Myr, for  $\mathfrak{M} = 4, 8$ from 1.5 to 2.5\,Myr.}
    \label{fig:outflows-mach}
\end{figure*}

Apart from that, we detect different outflow properties depending on the turbulent energy. We show radial velocity distributions in boxes with and without a jet in Fig.~\ref{fig:outflows-mach}. The distributions of radial velocity in boxes without jets have different widths due to the higher velocities associated with higher Mach numbers.

Fig.~\ref{fig:outflows-mach} shows that for the higher Mach number $\mathfrak{M} = 8$ (right panel) we see the strongest outflows in the hot phase whilst warm phase gas radial velocity is unaffected compared to the turbulent box with no jet. There was no hot gas in the turbulent box with lower Mach number $\mathfrak{M} = 2$ (left panel) prior to jet launching, but the jet drives the hot gas outflows up to ~250\,km\,s$^{-1}$. We also detect small outflows in the warm phase up to ~150\,km\,s$^{-1}$.

\section{Discussion} \label{sec:discussion}

In this section, we compare the results of our simulations with observational data and existing simulations.

\subsection{Comparison with observational data}

All of our runs including turbulence, for all values of jet power and turbulent Mach number probed, produce asymmetric jets. Such asymmetries are frequently observed in radio-loud \citep{Baldi2015} and radio-quiet \citep{Kharb2019} AGN. Sometimes these asymmetries are explained by the jet duty cycle \citep{Saikia2009}. However, we show that the turbulent ISM can be solely responsible for the deviations from straight collimated jets. Jets can be bent or appear as bubbles due to the dense gas walls and filaments that low-power jets cannot penetrate (see Fig.~\ref{fig:ram-pressure}, \ref{fig:ram-pressure-mach}).

Galactic scale outflows are one of the important features of AGN activity. AGN-driven outflows can be detected by measuring gas velocities using double-peaked emission lines. \cite{Kharb2019, Kharb2021} showed that around radio-quiet small-scale jets, there are outflows with velocities lower than 300\,km\,s$^{-1}$.  \cite{Roy2021} released the list of sources called red geysers where outflows were observed even though AGN activity was compact or unresolved. The typical velocities for red geyser outflows are also $\sim\!300$\,km\,s$^{-1}$. In our work, we measure a range of outflow velocities depending on the jet power and turbulent structure. In some cases, for example with Mach number $\mathfrak{M} = 4$ and intermediate jet power of $10^{40}$\,erg\,s$^{-1}$ (the middle panel of Fig.~\ref{fig:outflows}), we also find outflows with velocities that have the same order of magnitude --- 300-400\,km\,s$^{-1}$.

Moreover, x-ray observations of AGN in systems smaller than clusters suggest an abundance of hot gas perpendicular to the jet direction \citep{Fabbiano2022}. In Fig.~\ref{fig:jet-propagation}, we show that turbulence can cause lateral expansion of the hot jet material.

As a result, we show that the ISM can have a huge impact on the jet propagation and our simulations offer a simple explanation for many of the observed signatures of small scale jets in galaxies.

\subsection{Comparison with existing simulations}

Although details of jet launching implementation may vary in other simulations \citep{Sutherland2007, Wagner2012, Mukherjee2016, Mukherjee2018, Meenakshi2022, Dutta2024}, the resulting jet is similar. However, the main aspect of our simulations that makes it different from previous works is the turbulent medium environment. We use an ISM prescription containing filaments rather than discrete clumps, as observed in our Galaxy \citep{Heywood2022}.

Some results shown in this work are in agreement with existing jet-ISM interaction models. 
We demonstrate that jet material propagates extensively in lateral directions, consistent with findings by \cite{Dutta2024}, who showed that the head of the jet has to cut through higher-density regions while the cocoon expands through low-density channels. Similarly, \cite{Meenakshi2022} observed lateral outflows in the case of higher-powered jets.
Moreover, we find that lower-power jets are trapped in the ISM for a longer time (see Fig.~\ref{fig:jet-propagation}) which aligns with the conclusions of \cite{Mukherjee2016}. Furthermore, Fig.~\ref{fig:jet-propagation} demonstrates that jets propagate more slowly in a realistic ISM compared to a uniform, dilute gas box, supporting the results of \cite{Sutherland2007, Dutta2024}.

Nevertheless, there are differences in jet behavior compared to previous work.
First, in our model, the ISM structure is different — our box consists of the walls and filaments of warm gas. Most of the previous work has used clumps and clouds of warm gas, which has led to a different morphology of a cocoon \citep{Mukherjee2016, Mukherjee2018, Dutta2024}. Also, the fact that we run turbulence continuously throughout the simulation allows the ISM to fully control jet propagation for low-power jets. With an absence of continuous turbulence driving, the initial velocity quickly decays with time over a few Myrs and the ISM structure becomes less realistic. Second, in our simulation, defining the jet's head and cocoon radius presents a challenge compared to previous works \citep{Mukherjee2016, Mukherjee2018, Dutta2024}, since our jet’s shape is asymmetric and influenced by the cavities' shape.

\subsection{Limitations and potential improvements}

Our simulations lack some prescriptions of non-thermal processes that we plan to add in future works. For instance, magnetic fields and cosmic rays can affect jet and ambient medium mixing, as well as provide additional pressure support.

Another significant limitation of our work is its lacking a cold molecular phase. Unfortunately, resolving three phases of the ISM is computationally expensive with a jet launched in it. We expect jets to propagate more easily if gas can cool below $10^4$\,K because the media will become less filamentary and more clumpy.  However, further studies are required to explore this in detail.

We also do not include the effects of gravity. Existing models that apply a gravitational potential face difficulties reaching equilibrium \citep{Mukherjee2016, Mukherjee2018}. Therefore, in such simulations jets can be driven only for a few million years before the system collapses. However, these physical effects can affect jet propagation, so we plan to include a galactic gravitational potential in future work.

Here, we have explored only a handful of jet powers and ISM Mach numbers and only a single average number density. Accounting for different ISM properties may be necessary to explain the breadth of observational data of different jet morphologies in various galactic environments. Although we only consider fixed jet powers in this work, in the future the jet model can be coupled with an accretion model \citep{Weinberger2023}.


\section{Conclusions} \label{sec:conclusions}

In this paper, we present simulations of jets propagating in a turbulent ISM. We explored both different jet powers and different turbulent ISM Mach numbers to show that the jet behavior depends not just on its power, but also on its surroundings.

We found that low-power jets $L_{\text{jet}} \leq 10^{40}$\,erg\,s$^{-1}$ can be stalled within the central kiloparsec by a turbulent medium. This stalling can last at least 5\,Myr, with occasional bubbles of hot jet material transported outwards by the turbulent motion. These bubbles form when jets with power $L_{\text{jet}} < 10^{43}$\,erg\,s$^{-1}$ cannot push through walls of dense gas and are therefore forced to fill the cavities created by the turbulence. 
The ISM structure can also impact the direction of low-power jet propagation, with bent jets seen frequently in our simulations. We attribute this behavior to the insufficient ram pressure of the jet material relative to the ram pressure of  the filaments in the background media.

We emphasize that jet propagation in turbulent media differs from that in uniform gas. Jet material reaches further distances in a multiphase gas than in the uniform case with the same mean number density. However, the jet is significantly influenced by the surrounding walls, so it does not propagate as quickly or as far as it would in a dilute medium. These results are consistent even if we have different pattern of the walls and filaments in the turbulent box (see appendix~\ref{sec:appendix-stochasticity}).

When jets are stalled in the central kiloparsec of a galaxy, the overpressured cocoon can be still separated from the injection region and form a bubble of hot gas. This bubble can escape as an outflow, with the outflow velocity depending on the jet power. We show that more powerful jets ($10^{43}$\,erg\,s$^{-1}$) drive outflows in both hot and warm phases. In this case, warm phase outflows reach velocities of 500\, km\,s$^{-1}$, and hot phase outflows can be accelerated up to 0.1\,$c$. Hot phase outflows driven by intermediate-power jets are slower, reaching 750\,km\,s$^{-1}$. Warm ouflows can be detected as well up to 250\,km\,s$^{-1}$. For the lowest power jets ($10^{38}$\,erg\,s$^{-1}$) we do not detect any outflows driven outside of the central kiloparsec region. 

We launch an intermediate power jet  ($10^{40}$\,erg\,s$^{-1}$) into background media with varying Mach number. We detect the strongest hot gas outflows in media with stronger turbulence. Warm gas outflows are weaker, reaching $\approx 200$\,km\,s$^{-1}$ or becoming undetectable in highly turbulent media with $\mathfrak{M} = 8$.

We have taken the first steps towards characterising the interactions of AGN jets with the ISM of their host galaxy, and with further study of this parameter space will be in a position to develop the next-generation of sub-grid models for large-scale cosmological simulations.

\begin{acknowledgments}

We thank Mike Foley for providing us with the dimensionless turbulent box set up. We also thank Lisa Kewley, Ranieri Baldi, Pepi Fabbiano, Martin Elvis,  Grant Tremblay, and Charlie Conroy for helpful discussions about this work regarding AGN observations. We also thank Volker Springel, Max Grönke, Ralf Klessen, and Christoph Pfrommer for the useful insights about hydrodynamical simulations and a theoretical perspective of jet propagation.

This work was supported by the Simons Collaboration on ‘Learning the Universe’. The computations in this paper were run on the FASRC Cannon cluster supported by the FAS Division of Science Research Computing Group at Harvard University. GLB acknowledges support from the NSF (AST-2108470, AST-2307419 XSEDE grant MCA06N030), NASA TCAN award 80NSSC21K1053, and the Simons Foundation (grant 822237).

\end{acknowledgments}

%



\software{Arepo \citep{Springel2010},
          Gaepsi \citep{Gaepsi2011}}



\bibliographystyle{aasjournal}
\bibliography{jet-ism}


\appendix

\section{Test with higher resolution} 
\label{sec:appendix-resolution}

\begin{figure*}[tbh!]
    \centering
    \includegraphics[width=0.6\textwidth]{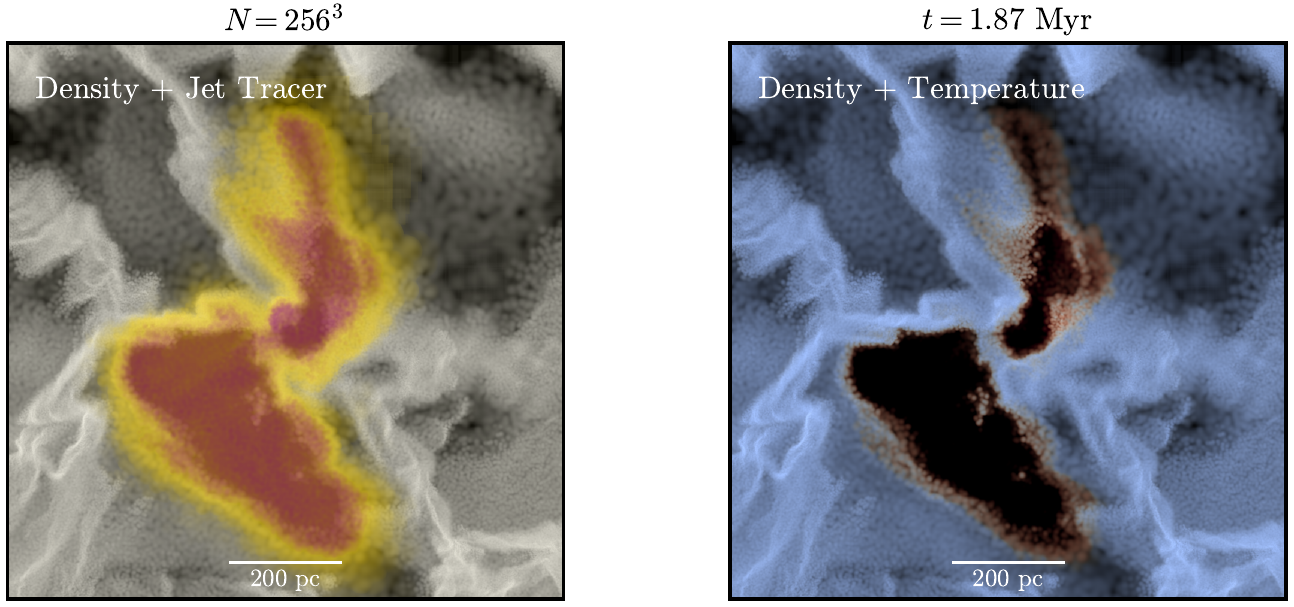}
    \includegraphics[width=0.6\textwidth]{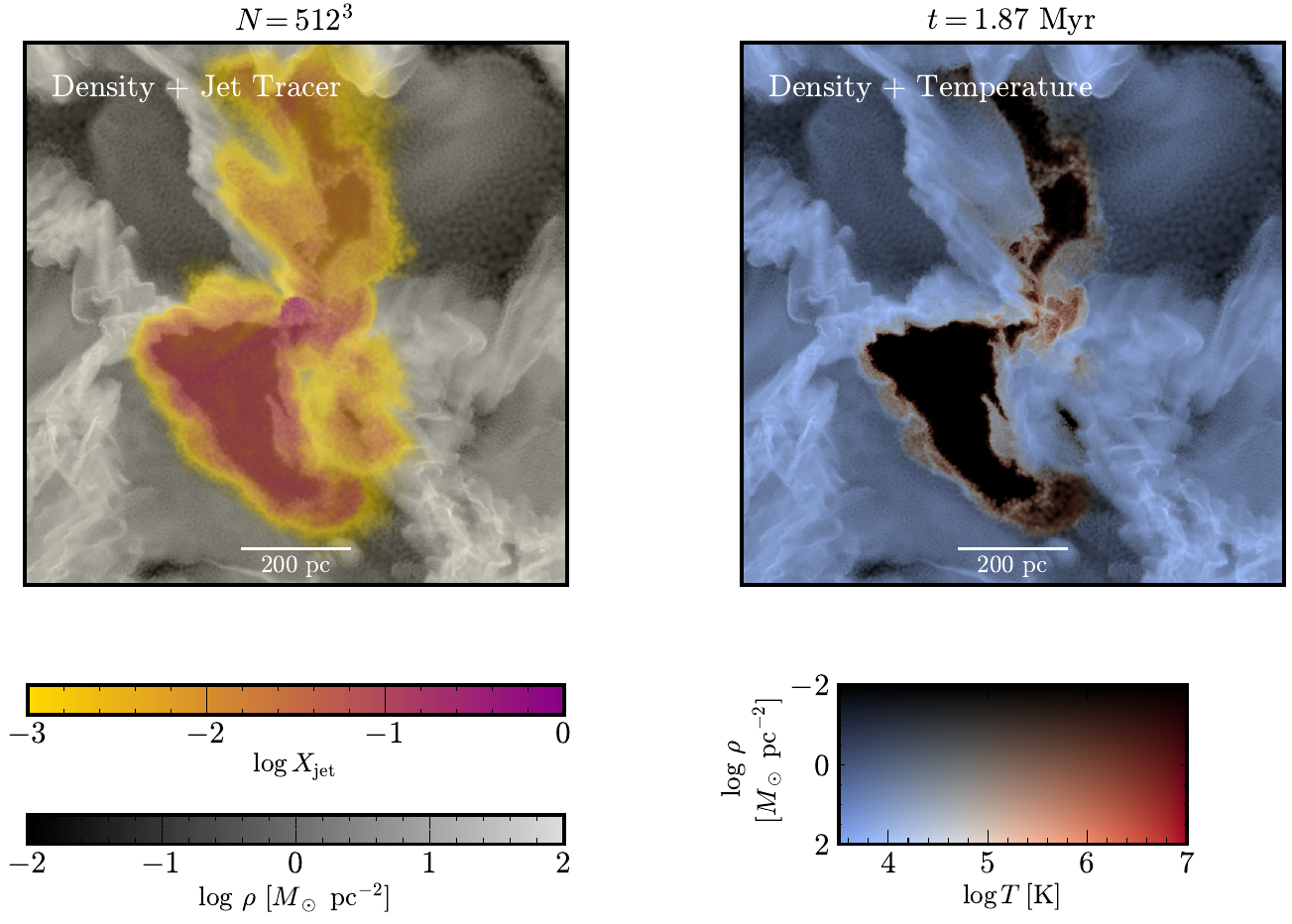}

    \caption{Projection map of jet propagation for jet power $10^{40}$\,erg\,s$^{-1}$ and $\mathfrak{M} = 4$ for runs with $N=256^3$ and $N=512^3$ in the top and bottom panels, respectively. Left panel shows black-white density colormap with overlay of jet tracer scalar. Right panel shows density map where colored by temperature value.}
    \label{fig:projection-maps-resolution}
\end{figure*}

Due to computational complexity, we limit the resolution of the simulations used throughout this paper to have a number of cells $N=256^3$. In order to test if resolution affects our results, we also run a simulation with 8 times the number of cells $N=512^3$. The jet propagation changes insignificantly and has the same properties as the runs with lower resolution (see Fig.~\ref{fig:projection-maps-resolution}).

\section{Test with different metallicity} 
\label{sec:appendix-metallicity}

\begin{figure*}[tbh!]
    \centering
    \includegraphics[width=0.5\textwidth]{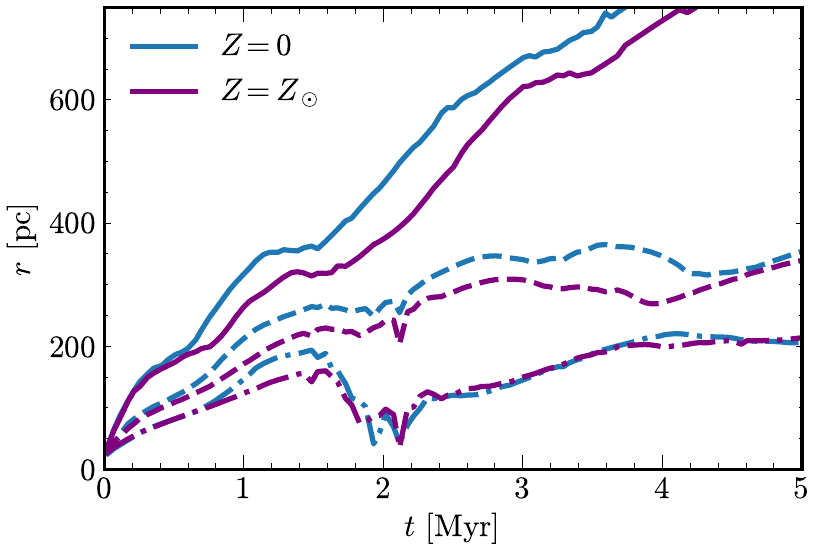}
    \caption{Evolution of the jet material distances $r_{50}, r_{80}$, and $r_{100}$ as a function of time for jet power $10^{40}$\,erg\,s$^{-1}$ in a turbulent box with Mach number $\mathfrak{M} = 4$. The solid, dashed, and dot-dashed lines correspond to $r_{100}$, $r_{80}$, and $r_{50}$, respectively. Blue and purple colors correspond to the metallicity $Z = 0$ and $Z = {\rm Z}_\odot$, respectively.}
    \label{fig:metallicity}
\end{figure*}

In this paper, we assume primordial abundance $Z = 0$. We perform the test where we use metallicity $Z =Z_\odot$. In Figure~\ref{fig:metallicity} we show jet propagation for the simulations with different metal abundances. We plot the radii of spheres containing 50\%, 80\%, and 100\% of the jet material, referred to as $r_{50}$, $r_{80}$, and $r_{100}$, respectively. We observe that jet material consistently reaches shorter distances when metal cooling is included, as indicated by the lower $r_{80}$ and $r_{100}$ lines in Fig.~\ref{fig:metallicity}. However, the difference is minimal, and the overall trend remains unaffected by variations in gas abundances.

\section{Jet in uniform ISM projection plot} 
\label{sec:appendix-uniform}

\begin{figure*}[tbh!]
    \centering
    \includegraphics[width=0.6\textwidth]{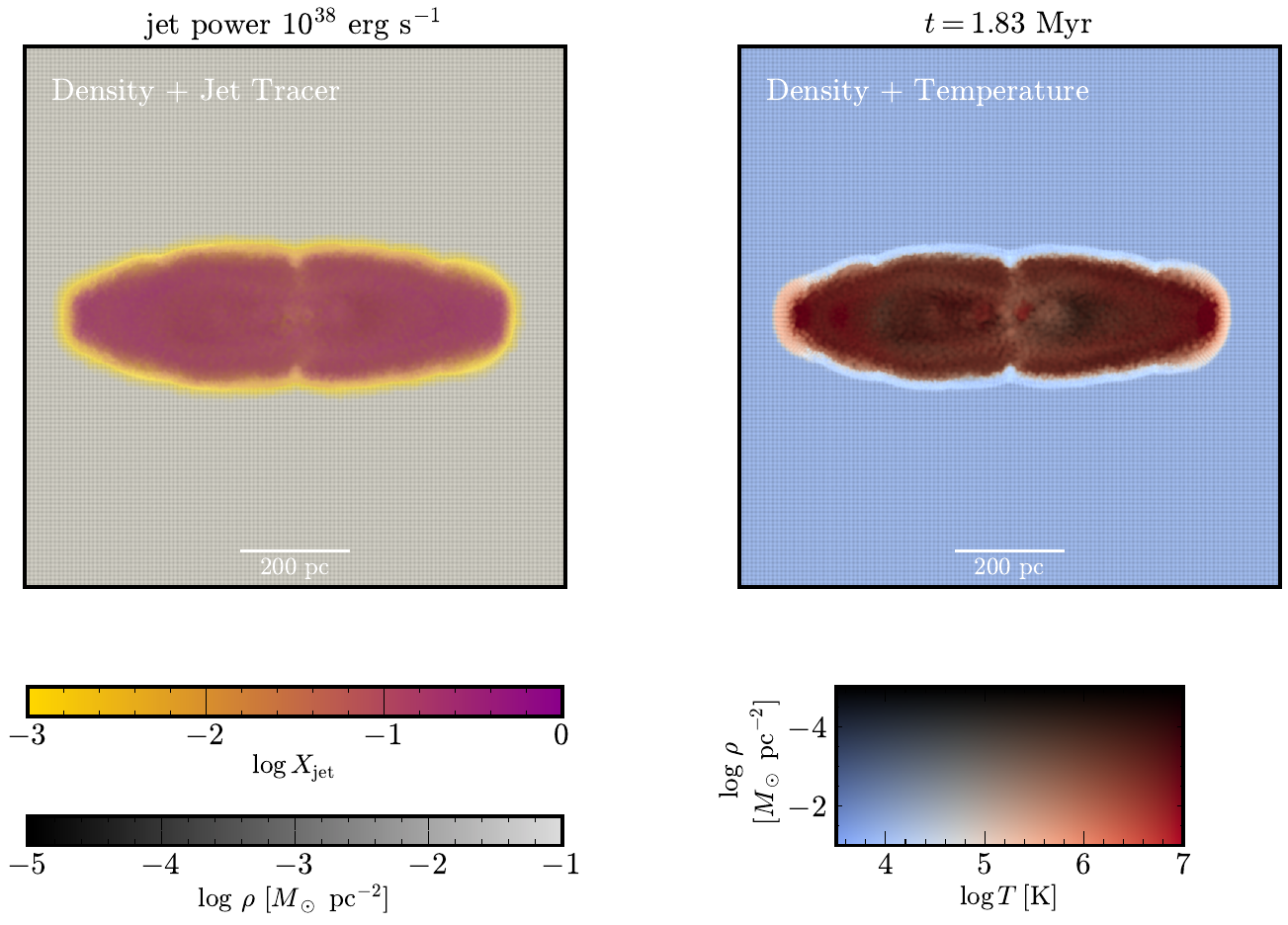}

    \caption{Jet propagation projection map for jet power $10^{38}$\,erg\,s$^{-1}$ in uniform medium with number density $n = 0.1$ cm$^{-3}$. Left panel shows black-white density colormap with overlay of jet tracer scalar. Right panel shows density map where colored by temperature value.}
    \label{fig:projection-map-uniform}
\end{figure*}

In this appendix, we show projection maps of a $10^{38}$\,erg\,s$^{-1}$ jet propagating in a uniform box (see Fig.~\ref{fig:projection-map-uniform}). The number density of the background medium  $n = 0.1$\,cm$^{-3}$ is low, therefore even such a weak jet can expand outside of the central kiloparsec. Also, the classic shape of a jet is reproduced --- with the head of a jet and cocoon. 

\section{Stochastic jet propagation} 
\label{sec:appendix-stochasticity}

We find that jets are highly affected by the turbulent ISM structure. For a jet power $10^{40}$\,erg\,s$^{-1}$, the majority of jet material is contained within the central kiloparsec, but turbulence can carry bubbles of jet material outside this region. We tested if this result can be replicated if the turbulent structure is different while keeping the same Mach number $\mathfrak{M} = 4$. To do this, we launched jets in turbulent boxes after different times 15, 17, 20, 22, and 25\,Myr, in order to test the effect of the different positions of walls and cavities.

\begin{figure*}[htp]
    \centering
    \includegraphics[width=0.5\textwidth]{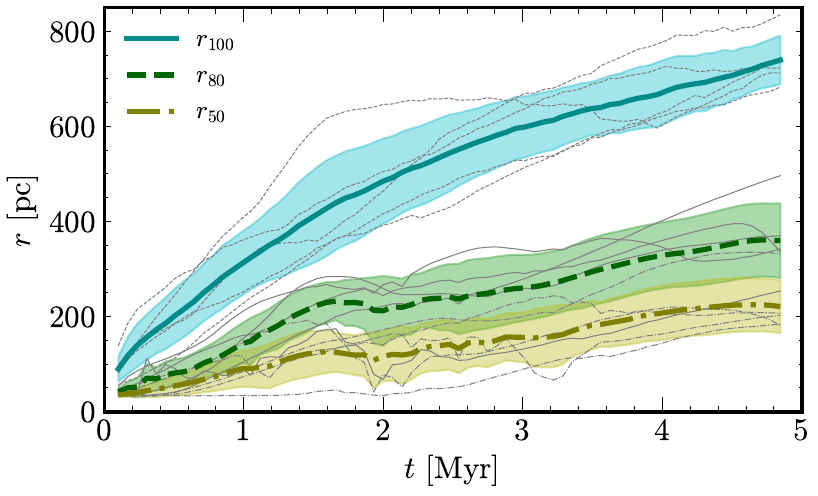}
    \caption{Evolution of the jet material distances $r_{50}, r_{80}$, and $r_{100}$ as a function of time for jet power $10^{40}$\,erg\,s$^{-1}$ in a turbulent box with Mach number $\mathfrak{M} = 4$. The solid blue line and filled region correspond to mean $r_{100}$ and its standard deviation, respectively. The dashed green line and filled region correspond to mean $r_{80}$ and its standard deviation, respectively. The dot-dashed olive line and filled region correspond to mean $r_{50}$ and its standard deviation, respectively. Gray lines with corresponding line styles represent different simulation realizations.}
    \label{fig:stochasticity}
\end{figure*}

In Figure~\ref{fig:stochasticity} we show the evolution of jet material distances for all of the runs starting at different times (grey lines), and the average and standard deviation of the runs (colored lines and area). Here we see that the jet has the same qualitative behavior regardless of the positions of walls (although the morphology of the jet structure varies). Different individual runs can differ from the average (i.e. the gray dashed line that goes above the blue dashed line before 3\,Myr), but overall they have the same trend and fluctuate around the average value. 

\section{Jet propagation in media with different Mach numbers} 
\label{sec:appendix-mach}

\begin{figure*}[htp]
    \hspace*{\fill}%
    \raisebox{-\height}{\includegraphics[width=0.3\textwidth]{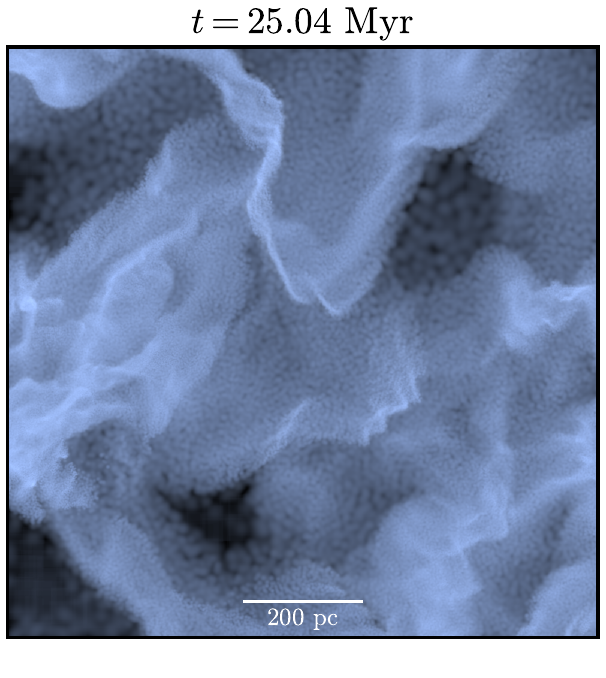}}%
    \hfill
    \raisebox{-\height}{\includegraphics[width=0.3\textwidth]{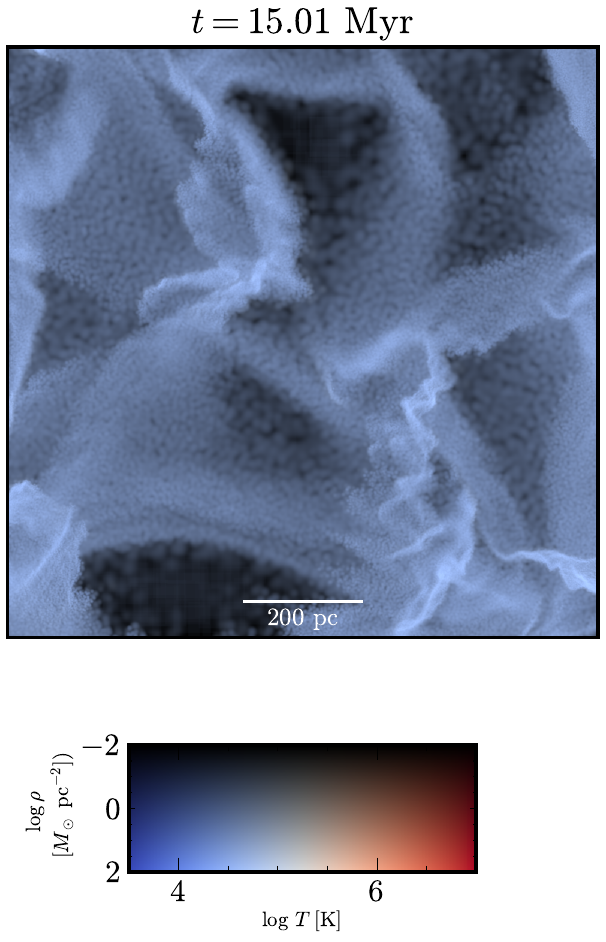}}%
    \hfill
    \raisebox{-\height}{\includegraphics[width=0.3\textwidth]{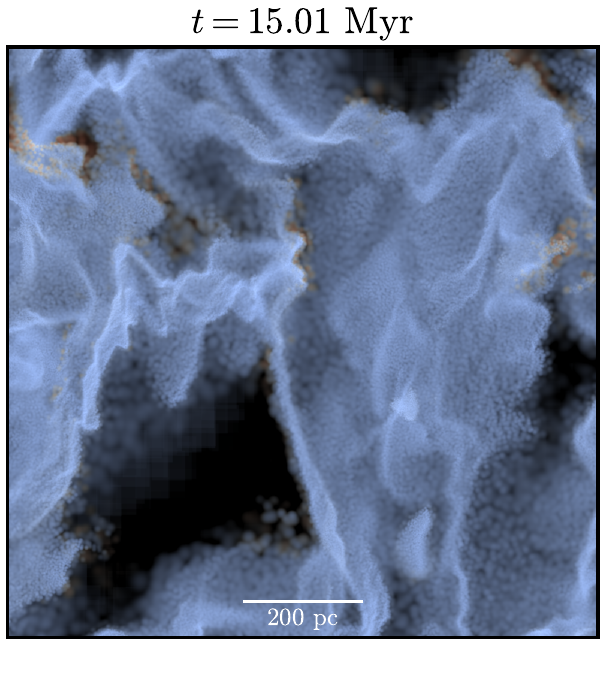}}%
    \caption{Projection maps of turbulent boxes with Mach numbers $\mathfrak{M} = 2, 4, 8$ in the left, middle, and right panels, respectively.}
    \label{fig:turb-machs}
\hspace*{\fill}
\end{figure*}

Different Mach number values indicate varying levels of turbulence strength, with higher Mach numbers corresponding to more turbulent gas. We use three Mach number values  $\mathfrak{M} = 2, 4, 8$ to analyze how jets propagate in different turbulent environments. In Fig.~\ref{fig:turb-machs} the three panels display the density structure in turbulent boxes with different Mach numbers. In the left panel, with $\mathfrak{M} = 2$ turbulence is the weakest, which is characterized by smoother filaments. For the highest Mach number, i.e. $\mathfrak{M} = 8$ in the right panel, hot shocks appear around the densest regions of the walls as they collide with each other.

\begin{figure*}[tbh!]
    \centering
    \includegraphics[width=0.6\textwidth]{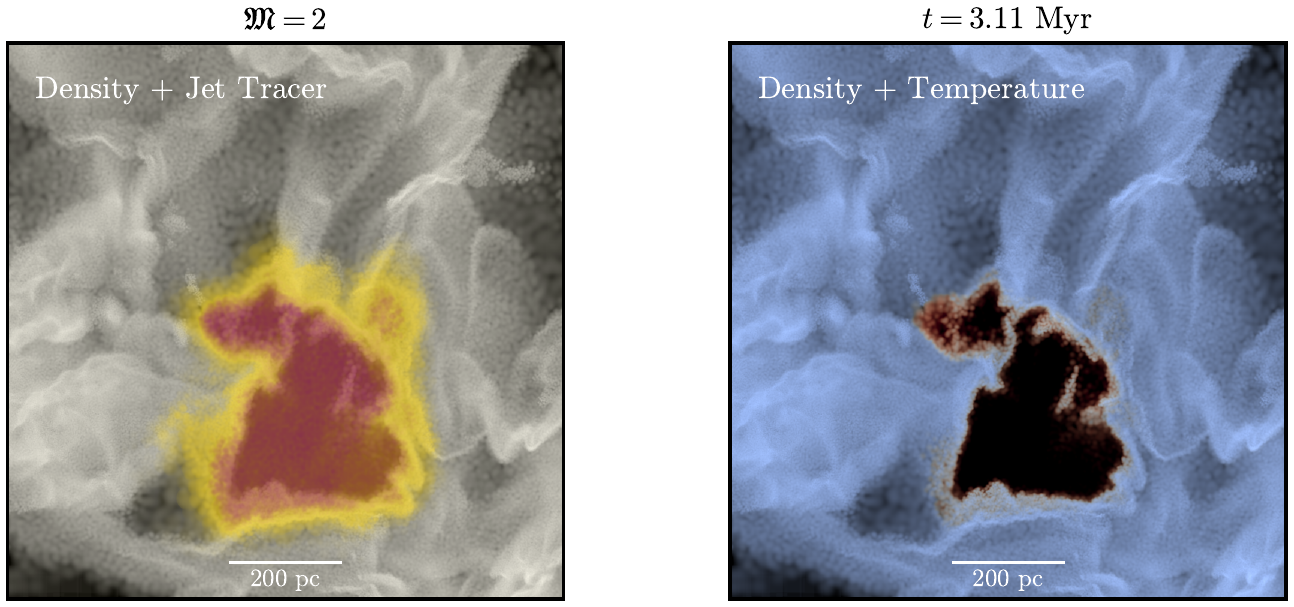}
    \includegraphics[width=0.6\textwidth]{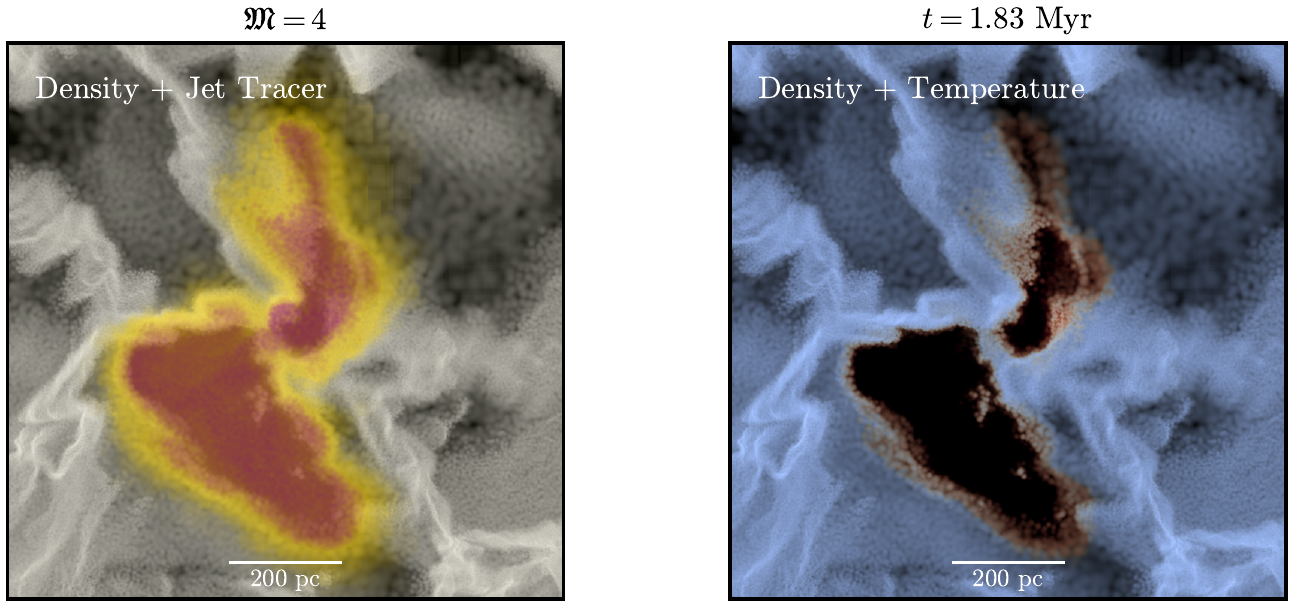}
    \includegraphics[width=0.6\textwidth]{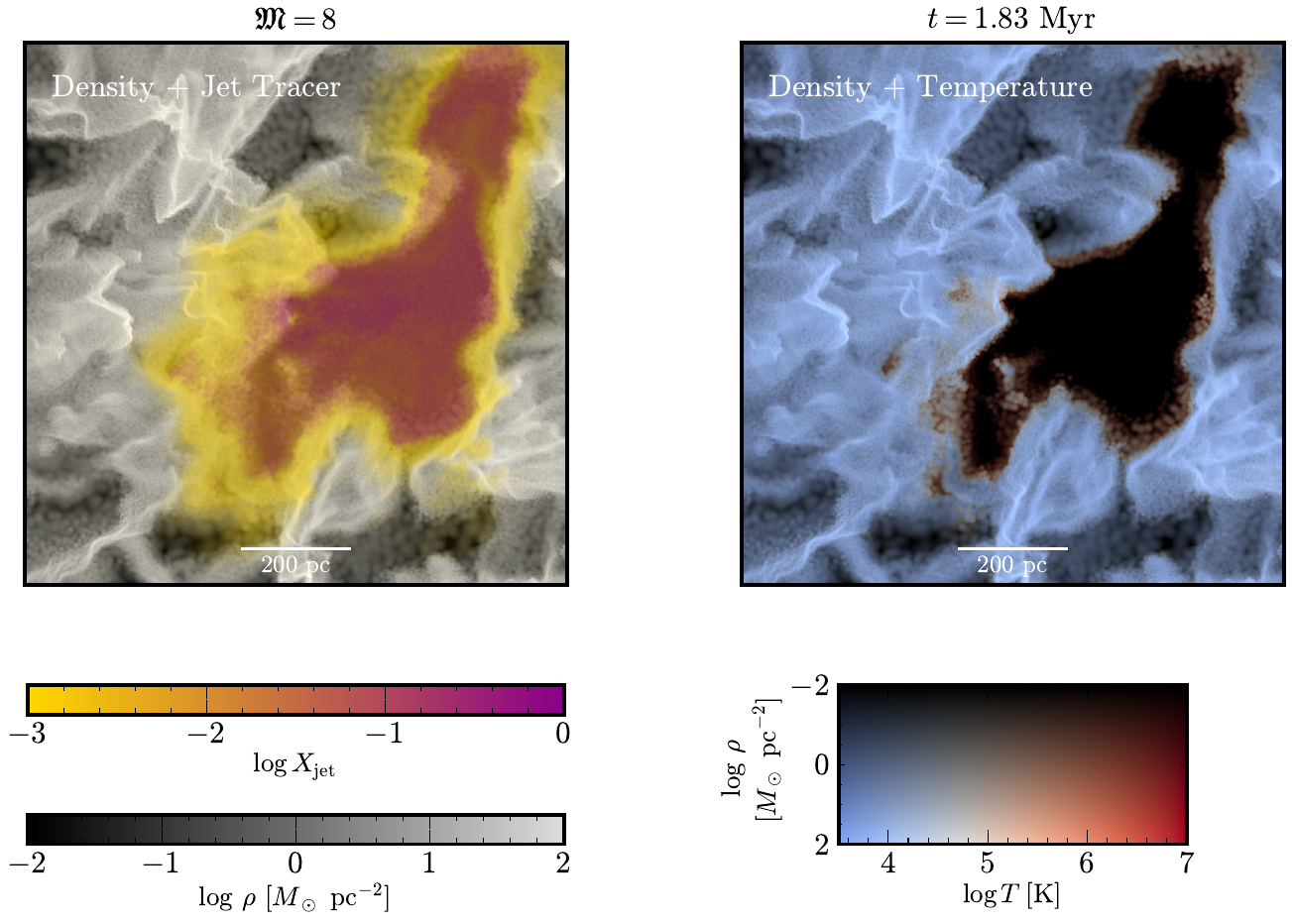}
    \caption{Projection map of jet propagation for jet power $10^{40}$\,erg\,s$^{-1}$ and $\mathfrak{M} = 2, 4, 8$  in the top, middle, and bottom panels, respectively. Left panel shows black-white density colormap with overlay of jet tracer scalar. Right panel shows density map where colored by temperature value}
    \label{fig:projection-maps-M}
\end{figure*}

In Fig.~\ref{fig:projection-maps-M} we show projection maps of jets propagating in media with different Mach numbers. In the top panel, the jet propagates in a medium with $\mathfrak{M} = 2$, approximately $3.6$\,Myr after the jet launch began. In the middle and bottom panels, we show jets propagating in media with $\mathfrak{M} = 4$ and $8$, respectively.  For higher Mach numbers, the jet propagates faster; therefore, we show it at approximately 1.8\,Myr after jet launch. In each case, we observe asymmetric jets that are perturbed by the background gas. For $\mathfrak{M} = 8$ (bottom panel in Fig.~\ref{fig:projection-maps-M}) we find more jet material in the warm phase than in less turbulent media.

\end{document}